\documentclass[%
twocolumn,
nofootinbib,
amsmath,amssymb,
aps
]{revtex4-1}
\usepackage{amsmath,amssymb}
\usepackage{graphicx}
\usepackage{dcolumn}
\usepackage{bm}

\usepackage{appendix}

\usepackage{qcircuit}
\usepackage{dsfont}
\usepackage{subcaption}
\usepackage{float} 
\usepackage{easyReview}

\begin{document}
	
	
	\title{Solving nonlinear differential equations on Quantum Computers: \\ A Fokker-Planck approach}
	\author{Felix Tennie}

	\email{f.tennie@imperial.ac.uk}

 	\affiliation{Imperial College London, Department of Aeronautics, Exhibition Road, London SW7 2BX, UK} 
  
	
	\author{Luca Magri}
 
 	\email{l.magri@imperial.ac.uk}
	\affiliation{Imperial College London, Department of Aeronautics, Exhibition Road, London SW7 2BX, UK} 

 	\affiliation{The Alan Turing Institute, London NW1 2DB, UK}

   	\affiliation{Politecnico di Torino, DIMEAS, Corso Duca degli Abruzzi, 24 10129 Torino, Italy}
  
	
	
	\date{\today}
	
	\begin{abstract}
	For quantum computers to become useful tools to physicists, engineers and computational scientists, quantum algorithms for solving nonlinear differential equations need to be developed. Despite recent advances, the quest for a solver that can integrate nonlinear dynamical systems with a quantum advantage, whilst being  realisable on available (or near-term) quantum hardware,  is an open challenge. In this paper, we propose to transform a nonlinear dynamical system into a linear system, which we integrate with quantum algorithms. Key to the method is the Fokker-Planck equation, which is a non-normal partial differential equation. Three integration strategies are proposed: (i) Forward-Euler stepping by unitary block encoding; (ii) Schr\"{o}dingerisation, and (iii) Forward-Euler stepping by linear addition of unitaries. We emulate the integration of prototypical nonlinear  systems with the proposed quantum solvers, and compare the output with the benchmark solutions of classical integrators. {We find that classical and quantum outputs are in good agreement.} This paper opens opportunities for solving nonlinear differential equations with quantum algorithms.
	\end{abstract}
	
	\maketitle
	
	
	\section{Introduction}
	
	Nonlinear differential equations are ubiquitous in Physics, Engineering, Chemistry, Biology, Economics, and various other subjects \cite[e.g.,][]{Jordan2007_nonlinear}. In many instances, computing numerical solutions on classical computers is challenging---and sometimes intractable, e.g., in turbulence---because of the large number of degrees of freedom \cite{Atkinson2009_NumericalSolutionDiffEqs}. Because of the exponential state space of quantum computers, and a theoretically proven quantum advantage of numerous \textit{linear} quantum algorithms, there is a growing interest in developing efficient quantum algorithms for integrating nonlinear dynamics  \cite{Lubasch2020,Kyriienko2021,Shukla2023,Lloyd2020,Liu2021,Krovi2023}. The state-of-the-art ans\"{a}tze can broadly be structured into three categories. 
	First, in hybrid quantum computing schemes \cite{Lubasch2020,Kyriienko2021,Shukla2023}, the integration of differential equations is formulated as a minimisation problem, which is solved by evaluating the loss function on a quantum computer with the parameters' updates being carried out on classical computers. These ans\"{a}tze can already be implemented on available quantum hardware; however, it is not yet clear whether they offer a scalable quantum advantage.
	Second, in a mean-field based ansatz \cite{Lloyd2020}, several copies of a quantum state evolve under a symmetric interaction Hamiltonian, which leads to a nonlinear evolution of individual copies. This solver offers a scalable quantum advantage, but it also requires extensive quantum hardware resources that are not yet available.
	{Third, nonlinear systems can be linearised via Carlemann embedding. This yields truncated equations, which are an approximation of the original nonlinear equations, which can be integrated with existing quantum algorithms \cite{Liu2021,Krovi2023}. }
 On the one hand, {for dissipative systems}, these ans\"{a}tze offer scaling advantages. On the other hand, the required quantum hardware resources are unlikely to be available in the near- to mid-term future.

	To integrate nonlinear dynamics on near- to mid-term available quantum hardware with a scalable quantum advantage,  new ans\"{a}tze and quantum solvers need to be developed.
	In this paper, we propose  quantum solvers that are based on the idea of transforming nonlinear differential equations into a corresponding Fokker-Planck equation\footnote{The Fokker-Planck equation governs the evolution of a probability distribution function of dynamical variables, which  are stochastically forced.}. 
Upon spatial discretisation, the Fokker-Planck equation becomes a master equation, which is a {\it linear} and non-normal Ordinary Differential Equation (ODE) system. 
Although linear ODE quantum solvers have been developed, their implementation on currently available quantum hardware is challenging. In this paper, we propose quantum solvers that have the potential to achieve integration of the Fokker-Planck equation on near- to mid-term available quantum hardware. 
 We propose three ways of integrating the master equation by taking into account its non-normal nature.  The computational output of this approach is a quantum state that is the time-evolved distribution function. From a dynamical systems' point of view, the state represents the solution of the dynamics and its uncertainty, which is of key interest to many disciplines, from weather and climate forecasting through turbulence,  to name a few. The proposed solvers are showcased by  numerically integrating  prototypical nonlinear systems. 
 %
 %

	The paper is organised as follows. In Sec.~\ref{sec:FokkerPlanck}, we review the Fokker-Planck formalism. In Sec.~\ref{sec:SpatialDiscretisationOfFP}, we discuss the spatial discretisation of the Fokker-Planck equation, and the properties of the resulting ODE system. The  non-normality of the Fokker-Planck operator, is discussed and tackled in Sec.~\ref{sec:QuantumODEsolversToFP}. In Sec.~\ref{sec:NearTermStrategies}, we discuss quantum solvers that prioritise structural simplicity over optimal asymptotic scaling. In particular, we develop a  solver that employs a linear combination of unitaries to achieve a Forward-Euler  integration. In Sec.~\ref{sec:IntegrationOfModelSystems}, we emulate these solvers for different model systems, thereby assessing the performance of the employed quantum solvers. We  discuss the results in Sec.~\ref{sec:discussion}, and conclude the paper in Sec.~\ref{sec:Outlook}.

	\section{Fokker-Planck Formalism}\label{sec:FokkerPlanck}

The evolution of nonlinear systems may be generally sensitive to the initial condition. Ensemble averages over several trajectories for varying initial conditions are a common tool to address this sensitivity. They can be formed using the distribution function of the dynamical variables. In this section, we revise the Fokker-Planck equation, which describes the evolution of the distribution function of the dynamical system's variables that are subject to stochastic forcing. Crucially, solving the Fokker-Planck equation\footnote{In the limit of zero diffusion.} is equivalent to integrating the nonlinear system \cite{risken2012fokker}, as discussed in this section. We consider a dynamical system governed by 
    \begin{equation}\label{eq:DynamilcalSystem01}
            \dot{\mathbf{x}} = \mathbf{f}(\mathbf{x},t) + \boldsymbol{\Gamma}(\mathbf{x},t),
    \end{equation}
where $\mathbf{x}(t) = (x_1,x_2,\ldots,x_d)^T$ denotes the state vector of $d$ dynamical variables; $\mathbf{f}$ may be a nonlinear function, and $\boldsymbol{\Gamma}$ is stochastic forcing, which models both the effect of the variables that are not modelled by $\mathbf{f}$ and aleatoric phenomena. Given an initial condition $\mathbf{x}(0) = \mathbf{x}_{in}$, the direct integration of the dynamical equations \eqref{eq:DynamilcalSystem01} yields a probabilistic outcome due to the stochastic forcings. Instead of computing a single trajectory of the system $\boldsymbol{\xi}_{\bar{\Gamma}}(t)$ for a specific realisation of the stochastic forcing $\bar{\Gamma}$, it is more informative to consider the evolution of a distribution function of initial values $\rho_{in} = \rho(\mathbf{x}_{in})$. In the limit of a precisely known initial condition, the initial distribution function becomes a $\delta$-function around $\mathbf{x}_{in}$, i.e.~$\rho_{in} = \delta(\mathbf{x}-\mathbf{x}_{in})$. The evolution of the distribution function $\rho$ subject to Eq.~\eqref{eq:DynamilcalSystem01} follows the Kramers-Moyal equation \cite{risken2012fokker}, which for a single dynamical variable (i.e., one degree of freedom, $d=1$) is
\begin{equation}\label{eq:KramerMoyal}
    \frac{\partial}{\partial t} \rho(x,t)= \sum_{k=1}^{\infty} \left(-\frac{\partial}{\partial x}\right)^k D^{[k]}(x,t)\rho(x,t).
\end{equation}
The coefficients $D^{[k]}$ can formally be written as expectation values with respect to the stochastic ensemble represented by $\Gamma$
\begin{equation}
    D^{[k]}(x,t) = \frac{1}{k!}\lim_{\tau \rightarrow 0} \frac{1}{\tau}\langle [\xi_{\Gamma}(t+\tau)-\xi_{\Gamma}(t)]^k\rangle _{|\xi_{\Gamma}(t) = x}, 
\end{equation}
{where $\langle\cdot\rangle$ is the expectation.} 
For systems with multiple dynamical variables $x_1,\ldots,x_d$, the Kramers-Moyal equation generalises to multiple index sums of tensorial coefficients $D^{[k_1,k_2,\ldots]}$. We refer the reader to Ref.~\cite{risken2012fokker} for more details.
%
By considering stochastic forcing $\boldsymbol{\Gamma}$ that {decomposes into} $\delta$-correlated Gaussian Langevin forcings, $\tilde{\Gamma}_{l}$, weighted by smooth functions $\boldsymbol{g}_l$,
\begin{align}
    &\boldsymbol{\Gamma}(\boldsymbol{x},t) = \sum_{l=1}^d \boldsymbol{g}_l(\mathbf{x},t) \tilde{\Gamma}_{l}(t), \label{eq:stochasticforces} \\
    \langle\tilde{\Gamma}_l \rangle &= 0, \quad \langle \tilde{\Gamma}_m(t)\tilde{\Gamma}_n(t')\rangle = 2 \delta_{mn} \delta(t-t'), 
\end{align} 
all terms of order $k\geq 3$  in the Kramers-Moyal equation vanish due to a corollary of Pawula's theorem  \cite{Pawula1967}.  
%
Thus, the Kramers-Moyal equation simplifies to the Fokker-Planck equation, which, for $d$ dynamical variables, is
    \begin{align}\label{eq:FokkerPlanck}
	 		\frac{\partial}{\partial t} \rho(\boldsymbol{x},t) = - &\sum_{i=1}^d \frac{\partial}{\partial x_i} \left( D^{(i)}(\boldsymbol{x},t) \rho(\boldsymbol{x},t)\right) \\ + &\sum_{i,j=1}^{d} \frac{\partial^2}{\partial x_i \partial x_j}\left(D^{(ij)}(\boldsymbol{x},t) \rho(\boldsymbol{x},t)\right) \nonumber \\ 
            & \equiv \hat{\mathcal{F}}\rho({\boldsymbol{x},t}) .\label{eq:DefinitionFokkerPlanckOperator}
 	\end{align}
The drifts $D^{(i)}$ and diffusion terms $D^{(ij)}$ form the linear Fokker-Planck operator, $\hat{\mathcal{F}}$, which can be computed as functions of the dynamics, $\mathbf{f} = (f_1,\ldots,f_d)^{T}$ \eqref{eq:DynamilcalSystem01}, and stochastic forcings \eqref{eq:stochasticforces} (cf.~Ref.~\cite{risken2012fokker} for more details)
    \begin{align}
    D^{(i)}(\boldsymbol{x},t) &= f_i(\boldsymbol{x},t) + \sum_{k,l=1}^d g_{kl}(\boldsymbol{x},t) \frac{\partial}{\partial x_k} g_{i l}(\boldsymbol{x},t) \label{eq:DefinitionDriftCoefficient}, \\
    D^{(ij)}(\boldsymbol{x},t) &=   \sum_{k}^{d} g_{ik}(\boldsymbol{x},t) g_{jk}(\boldsymbol{x},t).\label{eq:DefDiffusionCoefficient}
    \end{align}
The  \textit{nonlinear} dynamical system \eqref{eq:DynamilcalSystem01} has been transformed into a \textit{linear} dynamical system, as defined by the non-normal Fokker-Planck operator,  $\hat{\mathcal{F}}$.
When integrating the Fokker-Planck equation on a quantum computer, the problem specific drift and diffusion terms need to be readily computable and  accessible in form of oracles, which will be discussed in Sec.~\ref{sec:QuantumODEsolversToFP}.
For a vanishing forcing $\boldsymbol{\Gamma}=0$, the dynamical system in Eq.~\eqref{eq:DynamilcalSystem01} is \textit{deterministic}, thus, the Fokker-Planck equation simplifies to the Liouville equation, $\partial_t \rho + \boldsymbol{\nabla}\cdot (\boldsymbol{f} \rho)=0$. 
Although analytical solutions to the Liouville equation for an initial $\delta$-distribution are  equivalent to integrating the deterministic nonlinear dynamical system, $\dot{\mathbf{x}}=\mathbf{f}(\mathbf{x},t)$, numerically, an initial $\delta$-distribution does not remain localised, and the correct normalisation of the distribution function is not guaranteed. To overcome this problem, the diffusion terms in the Fokker-Plank equation enable numerical stability. By integrating the Fokker-Planck equation, we can integrate nonlinear dynamics on quantum computers.

\section{Numerical integration of the Fokker-Planck equation on Quantum Computers}\label{sec:DiscretisationFokkerPlanck}

A number of quantum algorithms for solving linear PDEs have been developed recently \citep[e.g.,][]{Childs2021}. Based on a discretisation of the spatial domain, these algorithms rely on subroutines such as the Quantum Linear System Algorithm (QLSA) (i.e.~the original Harrow–Hassidim–Lloyd algorithm \cite{Harrow2009} or refinements thereof such as Ref.~\cite{Childs2017}), which pose a  challenge to currently available quantum hardware.  We start by considering consistent spatial discretisations of the Fokker-Planck equation with the inclusion of boundary conditions (\ref{sec:SpatialDiscretisationOfFP}), which leads to a system of homogeneous linear ODEs with the output being a probability density function. We then discuss the applicability of existing quantum ODE solvers to this system, and the asymptotic scaling of those solvers in subsection \ref{sec:QuantumODEsolversToFP}. Because existing quantum ODE solvers make use of the QLSA, they may not offer an immediately feasible route for integrating the Fokker-Planck equation on near- to mid-term available quantum hardware. Instead, we improve strategies based on unitary block encoding and the recently proposed Schr\"{o}dingerisation method, and develop an integrator based on the linear addition of unitaries method in Sec.~\ref{sec:NearTermStrategies}.

\subsection{Spatial discretisation of the Fokker-Planck equation}\label{sec:SpatialDiscretisationOfFP}

Let the domain of interest be the interval $[-l,+ l]$. We  define a grid $\{x_k=-l + k \Delta x\}_{k = 0 \ldots N-1}$ of $N$ equally distanced grid points, where $\Delta x = 2l/(N-1)$. With the introduction of the grid, the continuous probability distribution function $\rho(x,t)$ can be approximated by an $N$-dimensional vector $\boldsymbol{\rho}=(\rho(x_0),\rho(x_1),\ldots)^T$. Similarly, we can choose to represent the probability density distribution by a vector $\boldsymbol{p} = (p_0,p_1, \ldots)^T$ of probabilities defined by $ p_k = \Delta x\rho(x_k)$, i.e.~$p_k$ represents the probability of finding the dynamical system variable $x$ at a value within $x_k\pm \Delta x/2$. In this paper, we use $\boldsymbol{p}$.  The equations governing the evolution of $\boldsymbol{p}$ can be derived by semi-discretising the Fokker-Planck operator $\hat{\mathcal{F}}$ \eqref{eq:DefinitionFokkerPlanckOperator}, which takes the form of a master equation \cite{Holubec2019}
\begin{equation}\label{eq:MasterEquation}
    \dot{\boldsymbol{p}}= \boldsymbol{R}\, \boldsymbol{p},
\end{equation}
whose formal solution is
\begin{equation}
 \boldsymbol{p}(t) = \mathcal{T}\mathrm{e}^{\int_0^t\boldsymbol{R}(t') dt'} \boldsymbol{p}(t=0),
\end{equation}
where $\mathcal{T}$ is the time-ordering operator.

The form of the matrix operator $\boldsymbol{R}$ may  not be unique. For any discretisation ansatz, Eq.~\eqref{eq:MasterEquation} needs to converge towards the continuous Fokker-Planck equation \eqref{eq:FokkerPlanck} for an increasing number of grid points $N$. In addition,  $\boldsymbol{p}(t)$ should remain a probability distribution, i.e.~with all $p_k\geq 0$ and constant normalisation $|\boldsymbol{p}|_1 = 1$ when periodic or reflecting boundary conditions are imposed. 
For non-boundary components $p_1,\ldots,p_{N-2}$ of $\boldsymbol{p}$, the following rate equations ensure probability normalisation~\cite{Holubec2019}
\begin{equation}\label{eq:RateEquation}
\dot{p}_k = r_{k-1,k}\: p_{k-1} + r_{k+1,k}\: p_{k+1} - (r_{k,k-1} + r_{k,k+1})\: p_k,
\end{equation}
where $r_{i,j}$ can be interpreted as the (positive) flow rate from grid point $i$ to $j$. The detailed expressions for $r$ are reported in App.~\ref{app:RateEquationCoefficients}.
%
Expressions for the matrix elements of $\boldsymbol{R}$ can also be derived by replacing the spatial derivative operators in the Fokker-Planck equation with finite differences. Using the central difference approximation,  we find for non-boundary grid points that
\begin{align}\label{eq:finiteDifferenceRateEquation}
\dot{p}_k = \frac{1}{\Delta x}\Bigg( \Big(\frac{1}{2}D^{(x)}_{k-1} + D^{(xx)}_{k-1}/\Delta x\Big)\: p_{k-1}  - (2D^{(xx)}_{k}/\Delta x)\: p_k\nonumber \\ 
+ \Big(-\frac{1}{2}D^{(x)}_{k+1} + D^{(xx)}_{k+1}/\Delta x\Big)\:p_{k+1} \Bigg),  
\end{align}
where $D^{(x)}_j $ and $D^{(xx)}_j$ are the drift and diffusion coefficients evaluated at $x_j$.
Because Eq.~\eqref{eq:finiteDifferenceRateEquation} can be derived by expanding the flow rates $r$ in Eq.~\eqref{eq:RateEquation} in orders of $\Delta x$,  the evolution of $\boldsymbol{p}$ under Eq.~\eqref{eq:finiteDifferenceRateEquation} leads to a consistent probability distribution provided that all components of $\boldsymbol{p}$ remain positive throughout the evolution. This requires that the prefactors  of the off-diagonal components, $p_{k-1}$ and $p_{k+1}$, in Eq.~~\eqref{eq:finiteDifferenceRateEquation} are positive, and it is necessary to choose a grid size smaller than
\begin{equation}\label{eq:MinMeshSize}
    \Delta x \leq \min_{k}\Bigg|\frac{2 D^{(xx)}_k}{D^{(x)}_k}\Bigg|.
\end{equation}
Consequently, there is a lower bound on the number of grid points given the size of a domain of interest. Although some discretisation schemes, e.g.~\cite{Holubec2019}, manage to avoid negative off-diagonal matrix elements of $\boldsymbol{R}$ even for grid sizes larger than the lower bound~\eqref{eq:MinMeshSize}, the physical interpretation remains: resolving the diffusion part of the Fokker-Planck equation imposes a constraint on the grid size. As shown in Sec.~\ref{sec:QuantumODEsolversToFP}, this lower bound impacts the scaling of a quantum solver. 
In the flow rate framework \eqref{eq:RateEquation}, boundary conditions need to be enforced. For reflecting boundary conditions, $r_{-1,0} = r_{0,-1}= 0$; whereas for periodic boundary conditions, {we have to include flow rates $r_{N,1}$ and $r_{1,N}$ between site $k=N$ and site $k=1$}. Sources and absorbing boundary conditions can be included by using an auxiliary site from or to which there is only a single-directional flow rate. 

In multi-variable systems, the spatial semi-discretisation of the Fokker-Planck equation leads to a master equation similar to Eq.~\eqref{eq:MasterEquation}. The vector $\boldsymbol{p}$ is now formed by elements with multiple indices, e.g.~for two dynamical variables $x_1 = x$ and $x_2 = y$,
\begin{equation}
    \boldsymbol{p} = (p_{k_x = 0, k_y = 0},p_{0,1},p_{0,2},\ldots,p_{0,N_x-1},p_{1,0},\ldots)^T.
\end{equation}
Correspondingly, the matrix\footnote{In this paper, matrices are shown in italic bold and upper-case text, vectors are shown in italic bold and lower-case text, and operators are shown in italic text.} $\boldsymbol{R}$ can be expressed as a sum of outer products
\begin{equation}\label{eq:MultiDimDiscretisedFokkerPlanckOperator}
    \boldsymbol{R} = \sum_{j=0}^{N_y-1} \boldsymbol{R}_x(y_j) \otimes \boldsymbol{E}^{(N_y)}_{jj}  +\sum_{i=0}^{N_x-1} \boldsymbol{E}^{(N_x)}_{ii} \otimes \boldsymbol{R}_y(x_i),
\end{equation}
where $N_x$ and $N_y$ are the number of grid points in the $x$ and $y$ dimensions, and $E^{(N)}_{kl}$ represents the $N$-dimensional square matrix with a single non-zero entry of $``1"$ at position $(k,l)$. The matrices $\boldsymbol{R}_x(y_j)$ and $\boldsymbol{R}_y(x_i)$ are formed according to Eq.~\eqref{eq:RateEquation} or \eqref{eq:finiteDifferenceRateEquation} by using drift and diffusion coefficients evaluated at constant $x_i$ and $y_j$, respectively. For non-diagonal diffusion coefficients $D^{(ij)}$ \eqref{eq:DefDiffusionCoefficient} additional cross-terms need to be taken into account. However, since the scope of this paper is to utilise the Fokker-Planck equation for effectively integrating nonlinear dynamics on quantum computers, we proceed by assuming that the diffusion matrix \eqref{eq:DefDiffusionCoefficient} is diagonal. Alternatively, it can be diagonalised by variable transformation. 
The tensor-product structure of $\boldsymbol{R}$ can naturally be mapped onto operator structures on multi-partite Hilbert spaces. This is advantageous for quantum solvers that require either a Hamiltonian simulation of $\boldsymbol{R}$, or its hermitian and anti-hermitian parts such as Schr\"{o}dingerisation (Sec.~\ref{sec:Schrodingerisation}), or the integration by linear combination of unitaries (Sec.~\ref{sec:LinearCombinationOfUnitaries}) techniques.

In summary, the continuous Fokker-Planck equation and boundary conditions can be consistently discretised in form of a master equation \eqref{eq:MasterEquation}, which is linear and non-normal. 

\subsection{Applying existing Quantum ODE solvers to the semi-discretised Fokker-Planck equation}\label{sec:QuantumODEsolversToFP}

Both the Fokker-Planck operator $\hat{\mathcal{F}}$ \eqref{eq:DefinitionFokkerPlanckOperator} and its discretised counterpart $\boldsymbol{R}$ \eqref{eq:MasterEquation} are sparse, semi-stable (i.e., the maximum real part of the spectrum is zero) non-normal operators. Suitable quantum solvers for the corresponding homogeneous ODE system have been constructed in Refs.~\cite{Berry2014,Berry2017,Childs2020,Krovi2023}, of which Krovi's algorithm offers a particularly promising asymptotic scaling. It requires access to $\boldsymbol{R}$ in form of oracles that reveal the column and row indices of the non-zero entries and a bit representation of their matrix value. 

For the Fokker-Planck operator associated with a single dynamical variable,  Eqs.~\eqref{eq:RateEquation} and \eqref{eq:finiteDifferenceRateEquation} show row and column sparsities of $s_r=s_c = 3$, respectively, with all non-zero elements located on the main diagonal and two neighbouring lower and upper diagonals. This allows for a computationally efficient implementation of index oracles. The generalisation to systems with $d$ dynamical variables is straightforward and yields a row and column sparsity of $s_r = s_c = 1+ 2d$ (from the structure of Eq.~\eqref{eq:MultiDimDiscretisedFokkerPlanckOperator}). 

To construct an oracle for the value of matrix elements of $\boldsymbol{R}$, we need to determine a method of reversibly evaluating a bit representation of either the transition rates $r_{k-1,k}$ \eqref{eq:RateEquation} or the drift and diffusion coefficients \eqref{eq:finiteDifferenceRateEquation}. Although the exponential function in the definition of the transition rates potentially leads to accuracy issues, thereby requiring a larger bit-depth, the evaluation of $D^{(i)}$ and $D^{(ij)}$ in the finite difference ansatz for $\boldsymbol{R}$ can be carried out using quantum networks for elementary arithmetic operations \cite{Vedral1996}. These operations have a  computational cost that is polynomial in the bit representation depth. Consequently, quantum networks for elementary arithmetic operations can be used to form computationally efficient oracles for the values of the non-vanishing matrix elements of $\boldsymbol{R}$ in our ansatz.

Equipped with efficient oracles, we can now analyse the asymptotic scaling and computational cost for integrating the semi-discretised Fokker-Planck equation. Let $\boldsymbol{p}(t) = \mathrm{e}^{\boldsymbol{R} t}\boldsymbol{p}_0$ denote the time evolved distribution  function and $|\boldsymbol{p}(t)\rangle$ a corresponding normalised quantum state. Let further  $b = \max_{t\in [0,T]}||\boldsymbol{p}(t)||_2 / ||\boldsymbol{p}(T)||_2 $, $\bar{N} = \max_{j = 1 \ldots d}( N_j)$ and  $C(\boldsymbol{R}) = \mathrm{sup}_{t\in [0,T]} || \mathrm{e}^{\boldsymbol{R} t}||_2$.  Based on Theorem 7 in Ref.~\cite{Krovi2023}, we find that the number of oracle calls required for generating an integration output within 2-norm $\epsilon$-error distance of $|\boldsymbol{p}(T)\rangle$ with success probability $\Omega(1)$ scales as
\begin{align}\label{eq:oracleCount}
    \mathcal{O}\Big(b T ||\boldsymbol{R}||_2 C(\boldsymbol{R}) \cdot&  \nonumber \\
    \cdot  \mathrm{poly}\big[d,\log(\bar{N})&,\log\Big(\frac{1}{\epsilon}\Big),\log\big(T ||\boldsymbol{R}||_2 C(\boldsymbol{R}) \big) \big] \Big).
\end{align}
The related gate complexity of the algorithm is larger by a factor of at most
\begin{equation}
	\mathcal{O}\Big( \mathrm{polylog}\big[\frac{1}{\epsilon}, T ||\boldsymbol{R}||_2 \big] \Big).
\end{equation}
As seen in Eq.~\eqref{eq:MinMeshSize}, resolving diffusion imposes a bound on the grid size. Thus the polynomial dependence on the logarithm of $\bar{N}$ can be recast into a scaling with $\mathrm{polylog}[\, l \times \max_{i,x\in [-l,l]} \frac{D^{(i)}(x)}{D^{(ii)}(x)}]$. Consequently, the impact of drift and diffusion on the overall computational cost is mainly given by the linear factor of $||\boldsymbol{R}||_2$ in \eqref{eq:oracleCount}.

Although the explicit dependence of the oracle count \eqref{eq:oracleCount} on the integration time $T$ is difficult to evaluate since $b$ and $C(\boldsymbol{R})$ implicitly depend on $T$, a worst-case lower bound can be established using an algorithm-agnostic result by An et al.~\cite{An2022}: provided that at least two eigenvalues of the coefficient matrix have different real parts, there exist initial conditions such that the overall computational overhead for any quantum ODE solver is exponential in $T$. Since the Fokker-Planck operator $\boldsymbol{R}$ has a spectrum of eigenvalues with different real parts \cite{risken2012fokker}, we conclude that the number of required oracle calls must in some cases grow exponentially with the integration time.

In summary, existing state-of-the-art quantum ODE solvers can be deployed to integrate the semi-discretised Fokker-Planck equation. Their scaling advantage is given by a polylogarithmic dependence on the number of grid points. However, for certain initial conditions quantum solvers will encounter a computational overhead, which grows exponentially with the integration time. In the next section, we propose ans\"{a}tze that are suitable near-term strategies.


\section{Near-term strategies}\label{sec:NearTermStrategies}

Existing quantum linear ODE solvers rely on subroutines such as the QLSA, which are challenging to implement on near- and mid-term quantum hardware. In this section, we  develop three alternative approaches. Two approaches are based on unitary block encoding \cite{Lloyd2021} (Sec.~\ref{sec:BlockEncodingSolver}) and quantum simulation techniques \cite{Jin2023} (Sec.~\ref{sec:Schrodingerisation}). The third approach uses linear combinations of unitaries (Sec.~\ref{sec:LinearCombinationOfUnitaries}). We discuss the suitability of the three approaches in the context of integrating the Fokker-Planck equation. All three methods aim to minimise quantum hardware requirements, whilst accepting a potential reduction in the asymptotic scaling efficiency.

\subsection{Forward-Euler integration by unitary block encoding}\label{sec:BlockEncodingSolver}

The semi-discretised Fokker-Planck equation \eqref{eq:MasterEquation} can be integrated using the finite difference forward-Euler method with time step size $\Delta t$. This requires repeated multiplications of the initial state vector $\boldsymbol{p}(0)$ by the matrix $\boldsymbol{A} =  \boldsymbol{1} + \Delta t \boldsymbol{R}$. Let $|\boldsymbol{p}(0)\rangle$ represent a normalised quantum state that is proportional to the initial state vector. Since $A$ is not a unitary operator, we have to resort to unitary block encoding \cite{Lloyd2021} for implementing a quantum matrix multiplication.\footnote{A matrix is said to be block-encoded in a unitary operator $U$ if it forms a block in the matrix representation of $U$.} By using an additional ancilla qubit $|q_a\rangle$, we can form the unitary operator
\begin{align}\label{eq:blockEncoding}
	\big(U_{\Delta t}\big)_{ij} =& \begin{pmatrix}
		\sqrt{\boldsymbol{1} - \boldsymbol{A}^\dagger \boldsymbol{A} }& & \boldsymbol{A}^\dagger \\
		\boldsymbol{A} & & -\sqrt{\boldsymbol{1} - \boldsymbol{A} \boldsymbol{A}^\dagger  }
	\end{pmatrix}_{ij}, \nonumber \\
U_{\Delta t} = & \,|0_a\rangle \langle 0_a| \otimes	\sqrt{\boldsymbol{1} - A^\dagger A } +  |0_a\rangle \langle1_a| \otimes	A^\dagger \nonumber \\   &+|1_a\rangle \langle 0_a| \otimes A -  |0_a\rangle \langle 0_a| \otimes	\sqrt{\boldsymbol{1} - A A^\dagger  }.
\end{align} 
Propagation of the initial state by a single Euler step is achieved by applying $U_{\Delta t}$ to $|0\rangle \otimes |\boldsymbol{p}(0)\rangle$ followed by a projective measurement of the ancilla qubit and a subsequent bit flip $X_a$.\footnote{In this paper, the operators $X$, $Y$ and $Z$ refer to the Pauli operators \cite{Nielson2010}.} With probability of success
\begin{equation}\label{eq:PsuccessBlockEncoding}
P_{suc} = 1 + \Delta t \langle  \boldsymbol{p}(0)|(\boldsymbol{R} + \boldsymbol{R}^\dagger ) |\boldsymbol{p}(0)\rangle + \mathcal{O}(\Delta t^2), 
\end{equation} 
the updated state becomes  $|0\rangle\otimes A |\boldsymbol{p}(0)\rangle$, which is proportional to $|\boldsymbol{p}(\Delta t)\rangle + \mathcal{O}(\Delta t^2)$. Since $\boldsymbol{R}$ is a transition matrix, it has eigenvalues with non-positive real parts \cite{Keitzer1972}. Therefore, for integration step sizes of
\begin{equation}\label{eq:IntegrationTimeStepSize}
	\Delta t \leq \frac{\lambda_{min}^*(-\boldsymbol{R} - \boldsymbol{R}^\dagger)}{\lambda_{max}(\boldsymbol{R}\boldsymbol{R}^\dagger)}
\end{equation}
we find that $A A^\dagger \leq \boldsymbol{1}$. Here, $\lambda_{min}^*()$ denotes the smallest non-zero eigenvalue and $\lambda_{max}$ represents the maximum eigenvalue. The choice of a suitable integration time step size guarantees that $U_{\Delta t}$ is well defined and that $P_{suc} \leq 1$. For initial states $|\boldsymbol{p}(0)\rangle$ that are not eigenvectors of $\boldsymbol{R}$ with corresponding eigenvalue $0$, $P_{suc}$ is strictly less than one. Therefore, the probability for successfully implementing $n$ Euler steps decreases as $\mathcal{O}((1-\mathcal{O} (\Delta t))^n)$, which imposes limitations on the total integration time $T=n\Delta t$. Unitary block-encoding is not a straightforward task because the form of $U_{\Delta t}$ in Eq.~\eqref{eq:blockEncoding} does not immediately reveal the Hamiltonian $H$, which is required to achieve a unitary evolution $U_{\Delta t} =  \mathrm{e}^{-i H \tau}$. In order to overcome this challenge, we  develop a method for implementing Euler steps based on a linear combination of unitaries (Sec.~\ref{sec:LinearCombinationOfUnitaries}).

\begin{figure*}
	\makebox[\textwidth]{
		\includegraphics[width=0.45\linewidth]{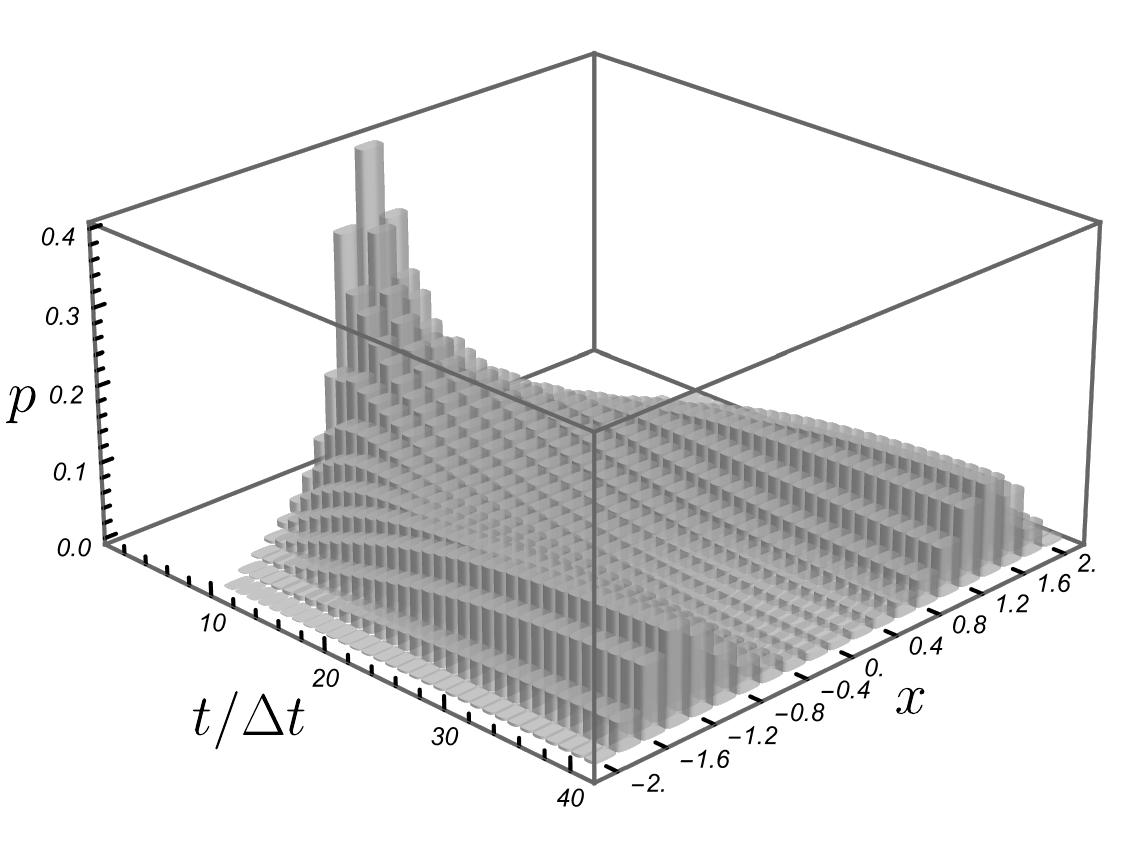}
		\includegraphics[width=0.45\linewidth]{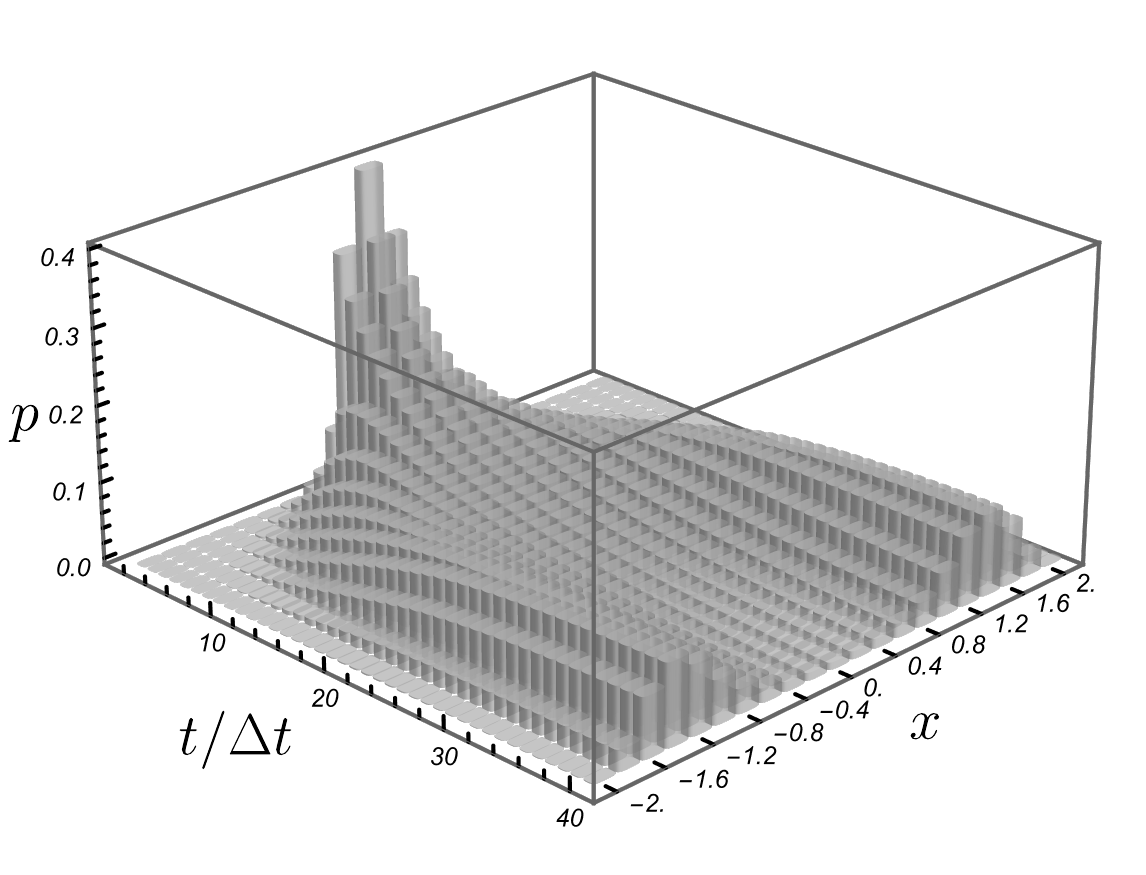}
	}
	\caption{Fokker-Planck evolution of an initially localised distribution subject to drift $f(x) = x -0.5 x^3$ and diffusion $D=0.15$. Left: emulated integration output of the Forward-Euler based quantum solver; right: emulated integration output of the Schr\"{o}dingerisation based quantum solver. For further details see Sec.~\ref{sec:IntegrationOfModelSystems}.}
	\label{fig:EvolutionDistributionFunction1dim}
\end{figure*}

\subsection{Integration by Schr\"{o}dingerisation}\label{sec:Schrodingerisation}

As shown in Sec.~\ref{sec:SpatialDiscretisationOfFP}, boundary conditions can readily be included in the semi-discretised master equation \eqref{eq:MasterEquation},  therefore, we derive its Schr\"{o}dingerisation~\cite{Jin2023} and prove that it is suitable for non-normal operators in Appendix~\ref{sec:ProofSchrodingerisation}. Let $\boldsymbol{R}_h = \frac{1}{2}(\boldsymbol{R} + \boldsymbol{R}^\dagger)$ and $\boldsymbol{R}_a = \frac{1}{2}(\boldsymbol{R} - \boldsymbol{R}^\dagger)$ represent the hermitian and anti-hermitian parts of $\boldsymbol{R}$. Let $\boldsymbol{p}(t) =  \mathrm{e}^{\boldsymbol{R} t} \boldsymbol{p}(0)$. Since $\boldsymbol{R}$ is not anti-hermitian, the propagator $ \mathrm{e}^{\boldsymbol{R} t}$ is not unitary, thus, it cannot  be realised by Hamiltonian simulation techniques. Instead, consider the function $\tilde{\boldsymbol{p}}(t,w) =  \mathrm{e}^{-|w|}\boldsymbol{p}(t)$. For a positive real number $w$, it suffices the partial differential equation
\begin{equation}\label{eq:NewSchroedingerisedPDE}
	\frac{d}{dt}\tilde{\boldsymbol{p}} = (- \boldsymbol{R}_h \partial_w + \boldsymbol{R}_a)\tilde{\boldsymbol{p}} .
\end{equation}
Given a solution of this differential equation for initial condition $\tilde{\boldsymbol{p}}_0 =  \mathrm{e}^{-|w|}\boldsymbol{p}(0)$, we can retrieve the solution to the original ODE by integration: $\boldsymbol{p}(t) = \int_0^\infty \tilde{\boldsymbol{p}}(t,w)dw$. In order to solve \eqref{eq:NewSchroedingerisedPDE}, we take the Fourier transform with respect to $w$, which yields
\begin{equation}\label{eq:NewSchroedingerisedPDEtransformed}
\frac{d}{dt}\bar{\boldsymbol{p}} =  (- 2\pi i \eta \boldsymbol{R}_h  + \boldsymbol{R}_a)\bar{\boldsymbol{p}},
\end{equation}
where $\bar{\boldsymbol{p}}(t,\eta) = \int_{-\infty}^{\infty}  \mathrm{e}^{-2\pi i w \eta }\tilde{\boldsymbol{p}}(t,w)dw$. For any $\eta\in\mathbb{R}$, Eq.~\eqref{eq:NewSchroedingerisedPDEtransformed} represents a Schr\"{o}dinger equation, which can be simulated using Hamiltonian simulation techniques. The outcome of such a simulation is inverse-Fourier transformed to yield the solution for Eq.~\eqref{eq:NewSchroedingerisedPDE}.
Let us summarise the steps for Schr\"{o}dingerisation:
\begin{enumerate}
	\item Given a linear ODE \eqref{eq:MasterEquation} and some initial condition $\boldsymbol{p}(0)$, initialise a quantum register in a state $|\boldsymbol{p}(0)\rangle$ and add a continuous variable register $|\psi_\eta\rangle$ initialised such that $\langle \eta |\psi_\eta\rangle = \frac{2}{1+4\pi^2 \eta^2}$, the Fourier transform of $ \mathrm{e}^{-|w|}$.
	\item Carry out Hamiltonian simulation for the Hamiltonian $H_S(\eta) =  2\pi  \eta \boldsymbol{R}_h  + i \boldsymbol{R}_a$ conditioned on $\eta$.
	\item Take the inverse Fourier transformation and retrieve $|\boldsymbol{p}(t)\rangle$ by taking the partial trace over the continuous variable register.
\end{enumerate}
Because of the normalisation of quantum states, Schr\"{o}dingerisation can only be applied to linear operators whose eigenvalues have non-positive real parts. This is the case for the Fokker-Planck operator as noted before \cite{Keitzer1972}. 

To execute Schr\"{o}dingerisation on quantum hardware, we can either   discretise the Fourier transformation and its corresponding register, or design new quantum hardware on which a Hamiltonian of the form $H_S$ can be generated. Although the latter approach is of potential interest to quantum hardware engineers, an efficient discrete Fourier transformation is provided in form of the Quantum Fourier Transformation (QFT) \cite{Nielson2010}. Replacing the continuous Fourier transformation by the QFT introduces truncation and aliasing errors, which limit the accuracy of the final solution. Establishing analytical bounds on these errors is beyond the scope of this paper. Therefore, we  assess the performance of  Schr\"{o}dingerisation method when used to integrate the Fokker-Planck equation for a range of models in  Sec.~\ref{sec:IntegrationOfModelSystems}.

\subsection{Integration by linear combination of unitaries}\label{sec:LinearCombinationOfUnitaries}

As discussed in Sec.~\ref{sec:BlockEncodingSolver}, forward-Euler integration can be achieved by unitary block encoding. However, for most operators block-encoding is not a straightforward task \cite{Camps2022}. In this section, we therefore develop a  method for implementing forward-Euler steps based on linear combinations of unitaries. We recall that the action of a linear combination of two unitary operators $U_1$ and $U_2$ on some quantum state $|\psi\rangle$ is achieved as follows. Let 
\begin{equation}
	B = \frac{1}{\sqrt{\alpha_1 + \alpha_2}}\begin{pmatrix}
		\sqrt{\alpha_1} & - \sqrt{\alpha_2} \\
		\sqrt{\alpha_2} & \sqrt{\alpha_1}
	\end{pmatrix}, \quad \alpha_1,\alpha_2\geq 0, 
\end{equation}
and add an ancilla qubit $|\phi_a\rangle$ to the quantum register and define the unitary $U_{com} = |0_a\rangle\langle 0_a|\otimes U_1 + |1_a\rangle\langle 1_a|\otimes U_2$. By applying $B^\dagger U_{com} B$ to the state $|0_a\rangle\otimes|\psi\rangle$ and carrying out a projective measurement on the ancilla qubit, upon outcome `0', the non-normalised quantum state is
\begin{equation}
	|\psi'\rangle = \frac{1}{\alpha_1 + \alpha_2}\big(\alpha_1 U_1 + \alpha_2 U_2\big)|\psi\rangle.
\end{equation}
The probability of success is given by the norm of $|\psi'\rangle$.  To execute a forward-Euler step of size $\Delta t$, we set $\alpha_1 = 1$, $\alpha_2 = \sqrt{\Delta t}$ and add another ancilla qubit $|\phi_{a'}\rangle$ such that we can choose to implement the unitaries
\begin{equation}
	U_1 = \begin{pmatrix}
	\boldsymbol{1} &0 \\ 0 & \boldsymbol{1} 
	\end{pmatrix}, \quad
U_2 =  \begin{pmatrix}
	0 & \boldsymbol{1} \\ \boldsymbol{1} & 0 
\end{pmatrix}  \mathrm{e}^{ - i H_R \sqrt{\Delta t}
}, 
\end{equation}
where the Hamiltonian $H_R$ is given by
\begin{equation}\label{eq:FokkerPlanckLinearCombUnitHamiltonian}
	H_R =  \begin{pmatrix}
		0 & -i \boldsymbol{R}^\dagger \\ i \boldsymbol{R} &0 
	\end{pmatrix}.
\end{equation}
By using a Taylor expansion of $U_2$ we find that the linear combination of $U_1$ and $U_2$ acting on $|0_{a'}\rangle\otimes|\boldsymbol{p}(0)\rangle$ to leading orders in $\sqrt{\Delta t}$ yields
\begin{equation}
|0_{a'}\rangle \otimes \big((1 + \Delta t \boldsymbol{R}) |\boldsymbol{p}(0)\rangle + \mathcal{O}(\Delta t^2) \big) + \sqrt{\Delta t}|1_{a'}\rangle\otimes |\boldsymbol{p}(0)\rangle.
\end{equation}
With probability $P_{a'}  = (1-\mathcal{O}( \Delta t))$, measurement and post-selection of the ancilla $a'$ now yields a quantum state that is proportional to the forward-Euler propagated initial state. The probability for successfully combining the two unitaries computes to $P_a = (1-\mathcal{O}(\sqrt{\Delta t}))$, therefore,  the overall probability scales as $P_{suc} =  (1-\mathcal{O}(\sqrt{\Delta t}))$. This can be quadratically improved using amplitude amplification. Although this scaling is less efficient than that of the unitary block encoding approach \eqref{eq:PsuccessBlockEncoding}, the implementation of the scheme is  simpler. In addition to single qubit operations on the ancilla qubits, we only require direct Hamiltonian simulation of the well-defined Hamiltonian $H_R$ for short times $\sqrt{\Delta t}$. This can be accomplished  by implementing a Hamiltonian of the form  $iX_{a'}\otimes(R - R^\dagger)$ and $Y_{a'}\otimes(R + R^\dagger)$. Because of the structure of $\boldsymbol{R}$ (cf.~Eq.~\eqref{eq:MultiDimDiscretisedFokkerPlanckOperator}), this type of Hamiltonian is   sparse.

\begin{figure}
	\centering
	\includegraphics[width=0.9\linewidth]{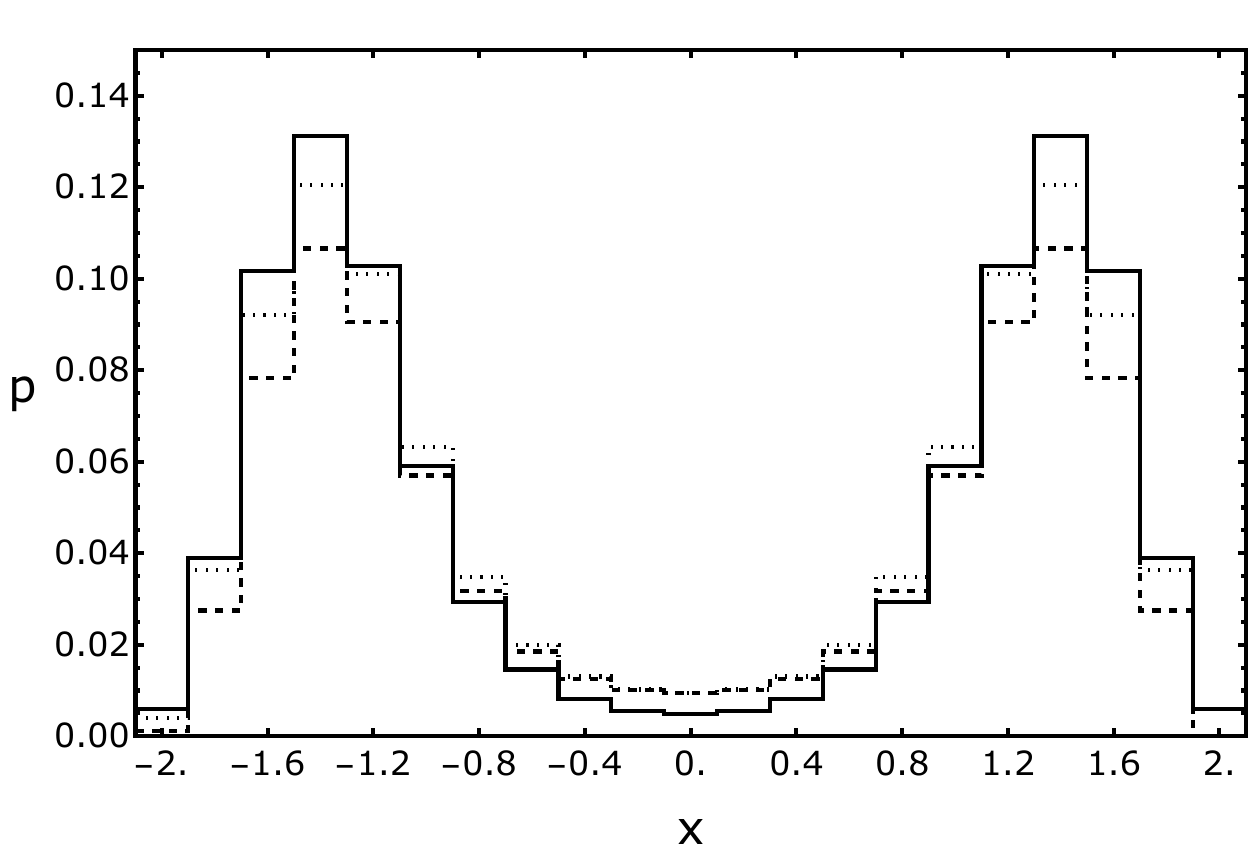}
	\caption{Long-term distribution functions computed by emulated quantum solver integration.  Forward-Euler (dotted line) and Schr\"{o}dingerisation solver outputs (dashed line) after $40$ integrations steps of size $\Delta t = 0.1$ for an initial distribution, localised around $x=0$ and evolving under Eq.~\eqref{eq:FokkerPlanck1Dmodel}, are shown. For comparison, the discretised distribution of the analytic steady-state  \eqref{eq:FokkerPlanck1DmodelSteadyStateDistribution} is displayed with a solid line. Despite minor discrepancies due to finite integration time, finite difference errors, and the constraints imposed by the reflecting boundary conditions, the quantum computation  well approximates  the analytical solution.}
	\label{fig:graphicfokkerpl1demulsteadystatecomparison}
\end{figure}

\section{Numerical results: Integration of prototypical nonlinear systems}\label{sec:IntegrationOfModelSystems}
We apply the quantum solver integration techniques \ref{sec:BlockEncodingSolver}, \ref{sec:Schrodingerisation} and \ref{sec:LinearCombinationOfUnitaries} to prototypical nonlinear models. First, we  consider a one-dimensional system $x(t)$ subject to the nonlinear dynamical equation (i.e., the drift), $\dot{x} = f(x) = x -\kappa x^3$ with $\kappa >0$. With a constant diffusion $D$, the Fokker-Planck equation \eqref{eq:FokkerPlanck} becomes 
\begin{equation}\label{eq:FokkerPlanck1Dmodel}
	\dot{\rho}(x,t) =  - \partial_x(f(x)\rho(x,t)) + D\partial_x^2\rho(x,t).
\end{equation}
In the limit of vanishing diffusion, all initial distributions converge to the fixed points of the drift $x_\pm = \pm 1/\sqrt{\kappa}$. For non-vanishing diffusion, we still observe a drift towards the fixed points, which  is ultimately balanced by diffusion when the system approaches the steady-state. The steady-state probability distribution can be computed analytically
\begin{equation}\label{eq:FokkerPlanck1DmodelSteadyStateDistribution}
	\rho_s(x) \propto  \mathrm{e}^{{\frac{1}{4D}(2x^2 - \kappa x^4)}}.
\end{equation}
Outside the interval $\mathcal{I} = [-1/\sqrt{\kappa},1/\sqrt{\kappa}]$ the distribution $\rho_s$ quickly decays to zero, thus, for an initial distribution localised within $\mathcal{I}$, we can select a finite domain $\tilde{\mathcal{I}}\supset \mathcal{I}$ and reflecting boundary conditions. 
\begin{figure*}
\makebox[\textwidth]{
	\includegraphics[width=0.45\linewidth]{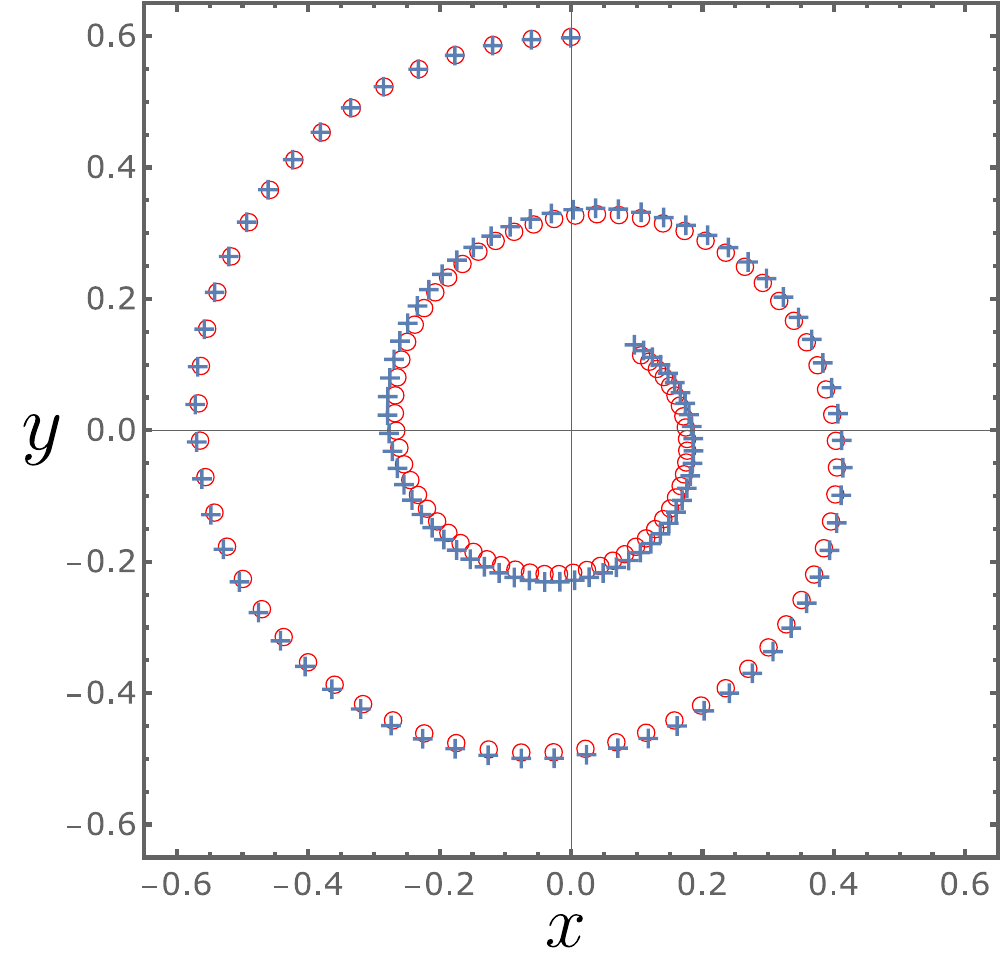}
	\includegraphics[width=0.45\linewidth]{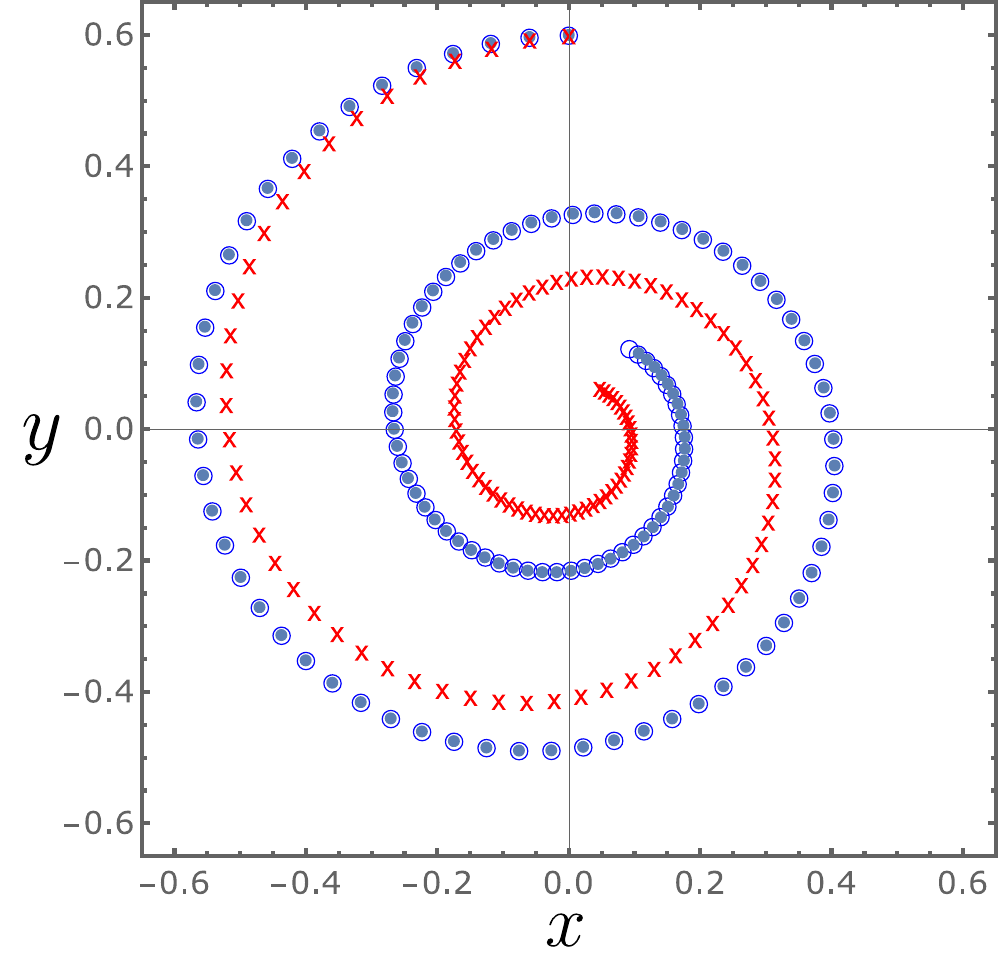}
	}
	\caption{Time evolution of the ensemble average position of a distribution $\rho(x,y)$ evolving under the Fokker-Planck equation related to Eq.~\eqref{eq:2dimSystemDrift} and constant diffusion. Left: Forward-Euler quantum solver solution (blue vertical crosses) vs classical output (red circles) for time step sizes $\Delta t = 0.01$. Right: Schr\"{o}dingerisation method output vs classical output (blue dots) for different discretisations of the Fourier register (blue circles correspond to $\eta \in [-10,10]$ and red diagonal crosses correspond to $\eta \in [-5,5]$ with $\Delta \eta = 0.01$).}
	\label{fig:EmulatedIntegrationOfInward2DSpiral}
\end{figure*}
We have emulated the integration of Eq.~\eqref{eq:FokkerPlanck1Dmodel} with the quantum solvers introduced in Sec.~\ref{sec:NearTermStrategies}. {For a fixed total integration time $T$, carrying out $n=T/\Delta t$ integration steps requires on average $E[T] = (p^{-n}-1)/(1-p)$ calls of a $p$-probabilistic quantum solver. Due to different success probabilities ($(1-\mathcal{O}({\Delta t})$ vs. $(1-\mathcal{O}(\sqrt{\Delta t})$ as shown in Secs.~\ref{sec:BlockEncodingSolver} and \ref{sec:LinearCombinationOfUnitaries}), both Forward-Euler based quantum solvers only differ by the average number of required measurement-and-post-selection processes. Therefore, we only display the output of one of the Forward-Euler solvers along with the output of the Schr\"{o}dingerisation based solver.}  For the emulated integration, we select a diffusion of $D=0.15$, $\kappa = 0.5$ and the domain $\tilde{\mathcal{I}}=[-2,2]$  with $\Delta x = 0.2$.\footnote{{While a high resolution of the spatial discretisation does not present a challenge to future quantum computers, limited classical computing resources impose contraints when emulating the integration of the quantum solvers. Our choice of the grid sizes takes this into account but suffices to reveal key features of the models and quantum solvers.}} The Forward-Euler based quantum solvers were emulated for $\Delta t = 0.1$. Results for this parameter choice are shown in Fig.~\ref{fig:EvolutionDistributionFunction1dim}. Further integration results for different parameter choices are reported in App.~\ref{app:AdditionalIntegrationOutcomes}.
The quantum solvers yield qualitatively equivalent outputs. The distribution function evolves towards local maxima around the fixed points and approaches the steady-state \eqref{eq:FokkerPlanck1DmodelSteadyStateDistribution}. The latter is explicitly demonstrated in Fig.~\ref{fig:graphicfokkerpl1demulsteadystatecomparison}, which shows a comparison between the distribution functions after $40$ integration steps and the expected discretised form of the analytic steady-state distribution.  The observed differences can be attributed to a finite integration time, finite difference approximation errors {and the constraints imposed by the reflecting boundary conditions.}

Next, we consider a two-dimensional nonlinear system subject to constant diffusion and a drift given by
\begin{equation}\label{eq:2dimSystemDrift}
	\begin{pmatrix}
		\dot{x} \\ \dot{y}
	\end{pmatrix}=\boldsymbol{f}_{2d}(x,y)=
\begin{pmatrix}
	x - \gamma x y^2 \\ -y  - \gamma x^2 y
\end{pmatrix}, \quad \gamma>0.
\end{equation}
The integral curves of the drift vector field $\boldsymbol{f}_{2d}$ are inward spirals, therefore, we can integrate the corresponding Fokker-Planck equation on a finite domain $\mathcal{I}_2$ imposing reflecting boundary conditions. Results of the emulated integration with the quantum solvers for $\gamma = 0.1$,  constant diffusion $D=0.15$ and a domain $\mathcal{I}_2=[-2,2]\times[-2,2]$ are shown in Fig.~\ref{fig:EmulatedIntegrationOfInward2DSpiral}. The distribution function $\rho(t,\boldsymbol{x})$ can be characterised by ensemble averages of the position vector $\langle \boldsymbol{x}\rangle = \int\boldsymbol{x}\rho(t,\boldsymbol{x})d\boldsymbol{x}$ and respective variances thereof. Results for different choices of the parameters are shown in App.~\ref{app:AdditionalIntegrationOutcomes}.

In order to gauge the performance of the quantum solvers, we have used classical numerical matrix exponentiation to compute the propagator $ \mathrm{e}^{\boldsymbol{R} t}$ of the master equation.
The left side of Fig.~\ref{fig:EmulatedIntegrationOfInward2DSpiral} shows a comparison between the classical solution and the output of the Forward-Euler quantum solver for time steps of size $\Delta t = 0.01$. The quantum solver successfully replicates the classical solution to a small numerical error. Results of the emulated integration by the Schr\"{o}dingerisation based quantum solver are shown on the right side of Fig.~\ref{fig:EmulatedIntegrationOfInward2DSpiral}. As discussed in Sec.~\ref{sec:Schrodingerisation}, a finite resolution of the Fourier register introduces truncation and aliasing errors. For a Fourier domain of $\eta \in [-10,10]$ with discretisation level $\Delta \eta = 0.01$, we find agreement between quantum and classical outputs (blue vertical crosses and dots). However, when truncating the inverse Fourier transformation to $\eta \in [-5,5]$, we find that the quantum output (red crosses)  diverges from the classical output. {Furthermore, aliasing errors have an even more drastic  impact on the quality of the solution. When decreasing the resolution of the Fourier register, we find that the quantum output diverges from the actual solution and forms a qualitatively different trajectory. A graphic display of these results is shown in Fig.~ Fig.~\ref{fig:graphicfokkerpl2demulschroedingerisationanddirectmatrexpo2} of App.~\ref{app:AdditionalIntegrationOutcomes}. This demonstrates that analytic bounds on the accuracy of the Schr\"{o}dingerisation method must be established in future work.}

\begin{figure}
	\centering
	\includegraphics[width=0.9\linewidth]{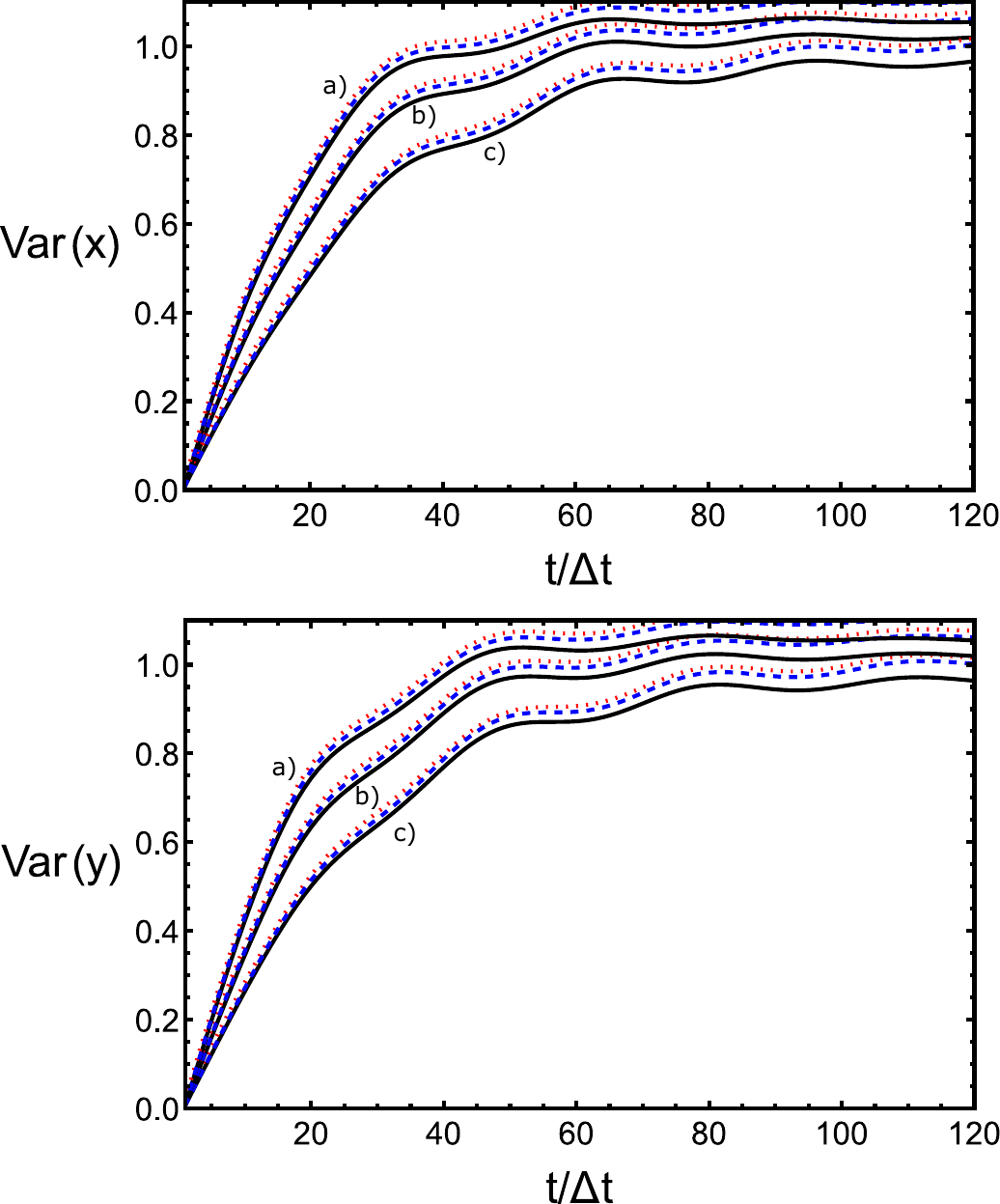}
	\caption[]{Spread of the distribution functions as measured by the variances of $x$ and $y$. The red dotted lines represent the classical output, whereas the blue dashed lines show the Forward-Euler quantum solver output. The solid lines mark the Schr\"{o}dingerisation method output based on a discrete Fourier register with $\eta\in[-10,10]$ and $\Delta \eta = 0.01$. $\gamma = 0.1$. The three groups of graphs belong to varying choices of the diffusion (a, $D=0.25$; b, $D=0.2$; c, $D = 0.15$).}
	\label{fig:graphicfokkerpl2demulvariances}
\end{figure}

Variances of the mean position represent a second key characteristic of localised distribution functions. Although  the mean of the distribution converges to the fixed point at the origin, diffusion  increases the spread of the distribution over time. A comparison of the classical and the quantum outputs is shown in Fig.~\ref{fig:graphicfokkerpl2demulvariances} for $\gamma =0.1$ and different values of the diffusion $D$. For both variances of the $x$- and $y$-position, the Forward-Euler quantum solver outputs (blue dashed) are almost identical with the classical output (red dotted). The Schr\"{o}dingerisation method output also replicates the classical solution with  small errors, which are attributed to truncation and aliasing errors.

\section{Discussion} \label{sec:discussion}
Similar to the emergence of classical GPUs and CPUs, future quantum hardware engineering should be customised to specific tasks to bridge the gap between quantum algorithm design and available quantum hardware resources. {In particular, the Hamiltonian, which is required for solving the Fokker-Planck equation on the newly developed quantum solver \ref{sec:LinearCombinationOfUnitaries}, can be written as a simple linear form of the discretised Fokker-Planck operator \eqref{eq:FokkerPlanckLinearCombUnitHamiltonian}.
For example, integration of a Fokker-Planck equation of $d=10$ dynamical variables, each spatially discretised by $1000\approx 2^{10}$ grid points, requires Hamiltonian simulation of $H_R$ \eqref{eq:FokkerPlanckLinearCombUnitHamiltonian}, which in this case becomes a sparse Hamiltonian on $101$ qubits. 
We conclude that near- to mid-term quantum computing technologies, which offer dozens of qubits with high-fidelity anywhere-to-anywhere connectivity, e.g.~ion traps or neutral atoms trapped in optical lattices, will be able to integrate Fokker-Planck equations of ten or more dynamical variables, a task that - if not intractable - requires extensive classical resources.}

\section{Conclusions}\label{sec:Outlook}

In this paper, we propose a strategy to solve nonlinear differential equations with quantum algorithms. The strategy is based on transforming the nonlinear dynamical system's equation into a linear equation via a Fokker-Planck approach. Upon spatial discretisation, the Fokker-Planck operator yields a master equation, which is a linear ordinary differential equation (ODE). 
First, we analyse the {computational complexity} of existing quantum linear solvers to integrate the master equations.
We find that the non-normal nature of the master equation poses a challenge. This is because both non-normality and the {the real part gap} in the spectrum of the Fokker-Planck operator can lead to an exponential computational overhead for certain initial conditions. Furthermore, existing quantum ODE solvers require quantum hardware resources that---albeit they might become available in the long-term future---are unavailable in the near-future. 
Second, we propose quantum solvers that aim to reduce the quantum hardware requirements for near-term applications, whilst accepting a potential reduction in the asymptotic scaling efficiency. The proposed quantum solvers are based on unitary block encoding, the Schr\"{o}dingerisation technique, and a linear addition of unitaries technique. We find that the unitary block encoding and the linear addition of unitaries techniques yield equivalent integration outputs up to leading order in the time step size. Both methods require measurement and post-selection steps and are therefore non-deterministic. {Since the respective probabilities are found to scale differently with the integration time step size, the expected number of solver calls will be larger for the linear addition of unitaries technique.}
Third, we emulate the integration of  prototypical nonlinear  models with the proposed quantum solvers. For various parameter settings, we find a good agreement between quantum and classical integration outputs for the unitary block encoding and the linear addition of unitaries techniques. The accuracy of the Schr\"{o}dingerisation technique is limited by truncation and aliasing errors arising from the discretisation of the Fourier register. Our results indicate that further research should be carried out to establish analytic bounds on these errors.

\section{Acknoweledgements}
Funding: The authors acknowledge financial support from the UKRI New Horizon grant EP/X017249/1. L.M. is grateful for the support from the ERC Starting Grant PhyCo 949388, and the grant EU-PNRR YoungResearcher TWIN ERC-PI\_0000005.

\appendix

\section{Rate equation coefficients}\label{app:RateEquationCoefficients}

In this appendix, we briefly present the explicit form of the transition rates as derived in Ref.~\cite{Holubec2019}. In a system of a single dynamical variable define the potential
\begin{equation}\label{eq:PseudoPotential}
V(x) = - \int  D^{(x)}(x')/ D^{(xx)}(x')dx'.
\end{equation}
Thermodynamically consistent transition rates are given by
\begin{equation}
	r_{k,k\pm 1} = \frac{D^{(xx)}(x_k)}{\Delta x^2}  \exp\left(-\frac{V(x_{k\pm 1})- V(x_k)}{2}\right).
\end{equation}
The generalisation to systems of $d$ dynamical variables is straightforward and requires the definition of $d$ pseudo-potentials similar to \eqref{eq:PsuccessBlockEncoding}. For positive diffusion the transition rates are necessarily positive. Furthermore,  by truncating the Taylor expansion of the exponential function to the first order, we retrieve the finite difference expression in Eq.~\eqref{eq:finiteDifferenceRateEquation}.

\section{Proof of the Schr\"{o}dingerisation method}\label{sec:ProofSchrodingerisation}

We provide a proof for the Schr\"{o}dingerisation method. More specifically, we show that the steps outlined in Sec.~\ref{sec:Schrodingerisation} give rise to a propagator, which is identical to the propagator associated with the original ODE $\boldsymbol{P}(t)=  \mathrm{e}^{\boldsymbol{R} t}$. First observe that $\boldsymbol{P}$ suffices the differential equation and the initial condition
\begin{equation}\label{eq:PropagatorDiffEq}
	\dot{\boldsymbol{P}} = \boldsymbol{R} \boldsymbol{P}, \quad \boldsymbol{P}(0) = \boldsymbol{1}.
\end{equation} 
Following the steps, the propagator $\boldsymbol{P}_S(t,s)$ combining all steps apart from the final integration over the continuous variable register is 
\begin{equation}\label{eq:Propagator}
	\boldsymbol{P}_S(t,s) = \int_{\mathbb{\boldsymbol{R}}}\frac{2}{1+4\pi^2\eta^2}  \mathrm{e}^{-2\pi i \eta(\boldsymbol{R}_h t - s) + \boldsymbol{R}_a t}d\eta .
\end{equation}
By taking the derivative with respect to time, we find
\begin{align}
	\dot{\boldsymbol{P}}_S(t,s) =& \int_{\mathbb{R}}\frac{2}{1+4\pi^2\eta^2} (-2\pi i \eta \boldsymbol{R}_h ) \mathrm{e}^{-2\pi i \eta(\boldsymbol{R}_h t - s) + \boldsymbol{R}_a  t}d\eta \nonumber \\  & + \int_{\mathbb{R}}\frac{2}{1+4\pi^2\eta^2} ( \boldsymbol{R}_a) \mathrm{e}^{-2\pi i \eta(\boldsymbol{R}_h t - s) +\boldsymbol{R}_a} d\eta \nonumber \\
	=& -2\pi i \eta \boldsymbol{R}_h \boldsymbol{P}_S(t,s) +  \boldsymbol{R}_a \boldsymbol{P}_S(t,s)\nonumber \\
	=& -\partial_s \boldsymbol{R}_h \boldsymbol{P}_S(t,s) +  \boldsymbol{R}_a \boldsymbol{P}_S(t,s). \label{eq:PSequation}
\end{align}
The final line represents a partial differential equation for $\boldsymbol{P}_S(t,s)$ with initial condition $\boldsymbol{P}_S(0,s) = \boldsymbol{1} \mathrm{e}^{-|s|}$. Observe that for positive $s$, $ \mathrm{e}^{-|s|}\boldsymbol{P}(t)$ is a solution to the differential equation \eqref{eq:PSequation} and therefore $\boldsymbol{P}_S(t,s) =  \mathrm{e}^{-|s|}\boldsymbol{P}(t)$ on the domain of positive $s$. Consequently, the final step in the Schr\"{o}dingerisation method yields the original propagator
\begin{equation}
	\boldsymbol{P}(t) = \int_0^\infty \boldsymbol{P}_S(t,s') ds',
\end{equation}
which completes the proof.

\section{Parameter studies of the prototypical nonlinear models}\label{app:AdditionalIntegrationOutcomes}

\begin{figure}
	\centering
	\includegraphics[width=0.9\linewidth]{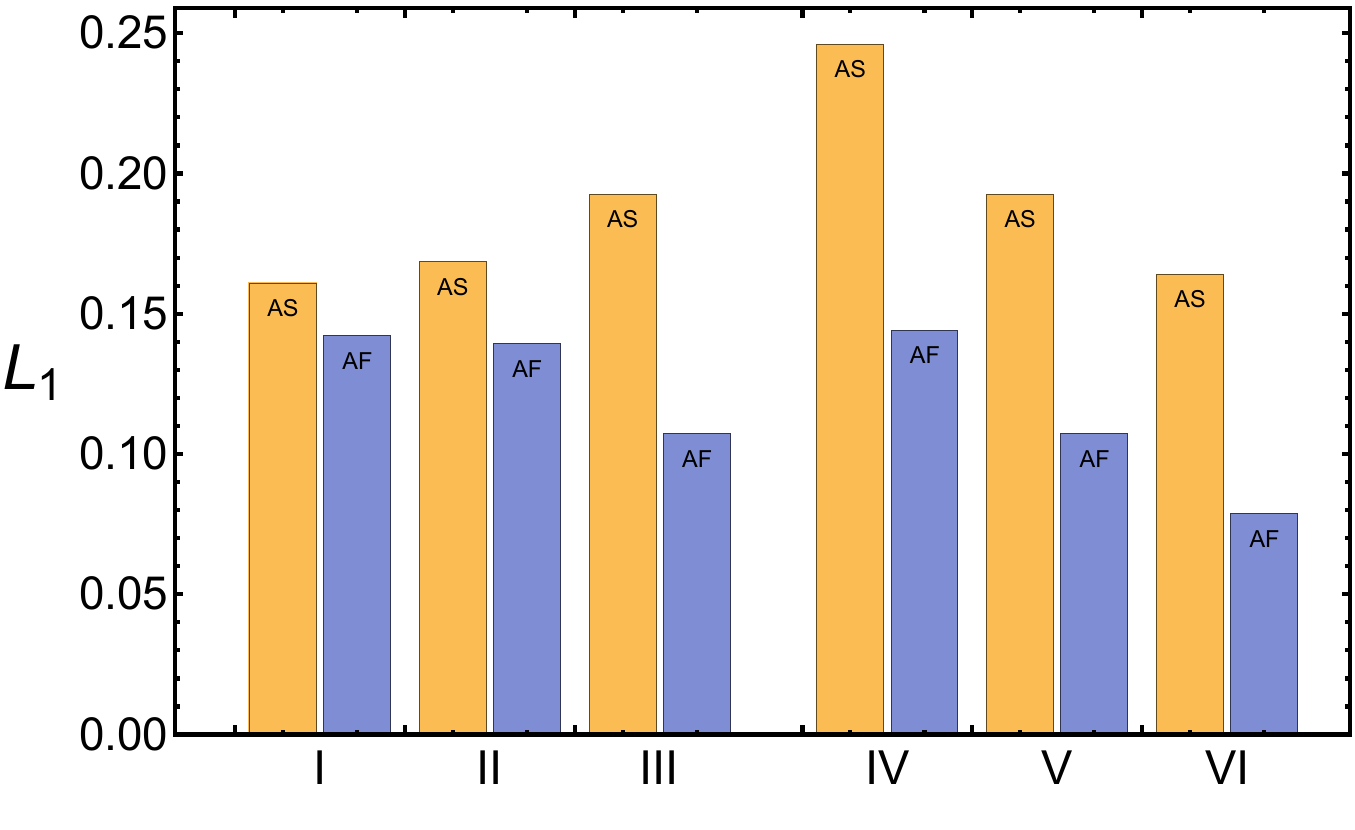}
	\caption{Trace distances $L_1$ between long-term distribution functions derived by emulated quantum solver integration and the analytic steady-state distribution \eqref{eq:FokkerPlanck1DmodelSteadyStateDistribution} after $40$ integrations steps of size $\Delta t = 0.1$ for an initial distribution, localised around $x=0$ and evolving under Eq.~\eqref{eq:FokkerPlanck1Dmodel}.  In each group, the left bar (AS) shows the distance between the analytic and the Schr\"{o}dingerisation solver outputs, and the right bar (AF) shows the distance between the analytic and the Forward-Euler quantum solver outputs. Groups I to III correspond to parameters $D=0.15$ and $\kappa = 0.3,0.4,0.5$, respectively.  Groups IV to VI correspond to parameters $\kappa = 0.5$ and $D = 0.12,0.15,0.18$, respectively.}
	\label{fig:graphicfokkerpl1dTraceDistances}
\end{figure}

\begin{figure*}
	\makebox[\textwidth]{
		\includegraphics[width=0.45\linewidth]{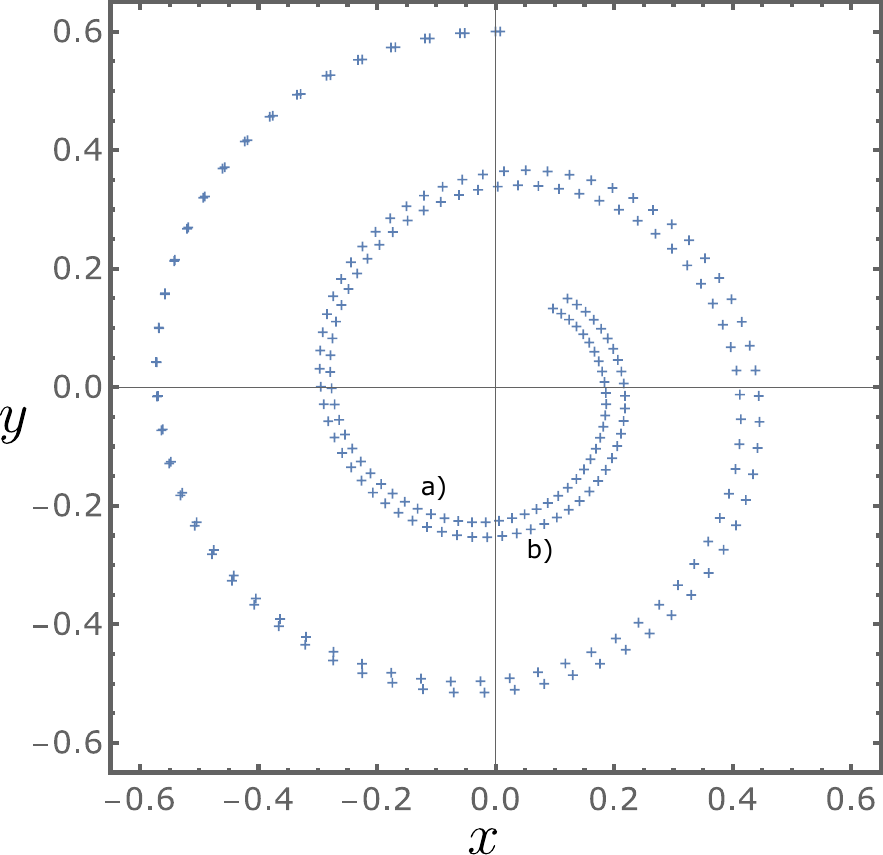}
		\includegraphics[width=0.44\linewidth]{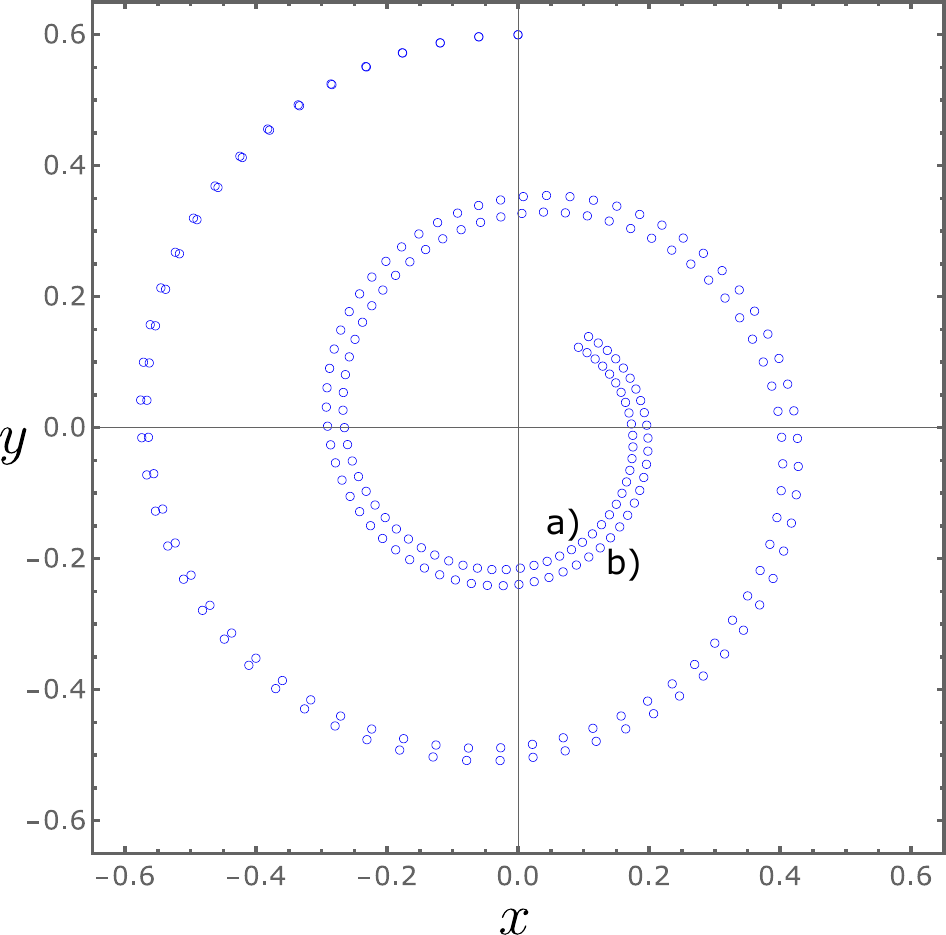}
	}
	\caption{Time evolution of the ensemble average position of a distribution $\rho(x,y)$ evolving under the Fokker-Planck equation related to Eq.~\eqref{eq:2dimSystemDrift} and constant diffusion. 
 Left: Forward-Euler quantum solver output for nonlinear coupling parameters $\kappa=0.1$ (a, inner spiral) and $\kappa = 0.06$ (b, outer spiral), diffusion of $D=0.15$ and integration time step sizes $\Delta t = 0.01$.
 Right: Schr\"{o}dingerisation method output for nonlinear coupling parameters $\kappa=0.1$ (a, inner spiral) and $\kappa = 0.06$ (b, outer spiral) and diffusion of $D=0.15$.}
\label{fig:EmulatedIntegrationOfInward2DSpiral_gamma}
\end{figure*}

In this appendix, we present the results of integrating the prototypical nonlinear models discussed in Sec.~\ref{sec:IntegrationOfModelSystems} for varying parameters. The one-dimensional system evolving under the Fokker-Planck equation \eqref{eq:FokkerPlanck1Dmodel} depends on the parameter $\kappa$ and the constant value of diffusion $D$. A comparison between different distribution functions is shown in Fig.~\ref{fig:graphicfokkerpl1dTraceDistances}. We choose the trace distance $L_1(\boldsymbol{p}^a,\boldsymbol{p}^b) = \sum{i}|p_i^a - p^b_i|$ between distributions $\boldsymbol{p}^a$ and $\boldsymbol{p}^b$, and compare the Schr\"{o}dingerisation solver and quantum Forward-Euler solver outputs after $40$ integration steps with the discretised analytic steady-state distribution \eqref{eq:FokkerPlanck1DmodelSteadyStateDistribution} for various parameter settings. 

The two-dimensional system \eqref{eq:2dimSystemDrift} is parametrised by $\gamma$. Integration outputs for variable $\gamma$ and constant diffusion are shown in Fig.~\ref{fig:EmulatedIntegrationOfInward2DSpiral_gamma}.

\begin{figure*}
	\centering
	\includegraphics[width=0.5\linewidth]{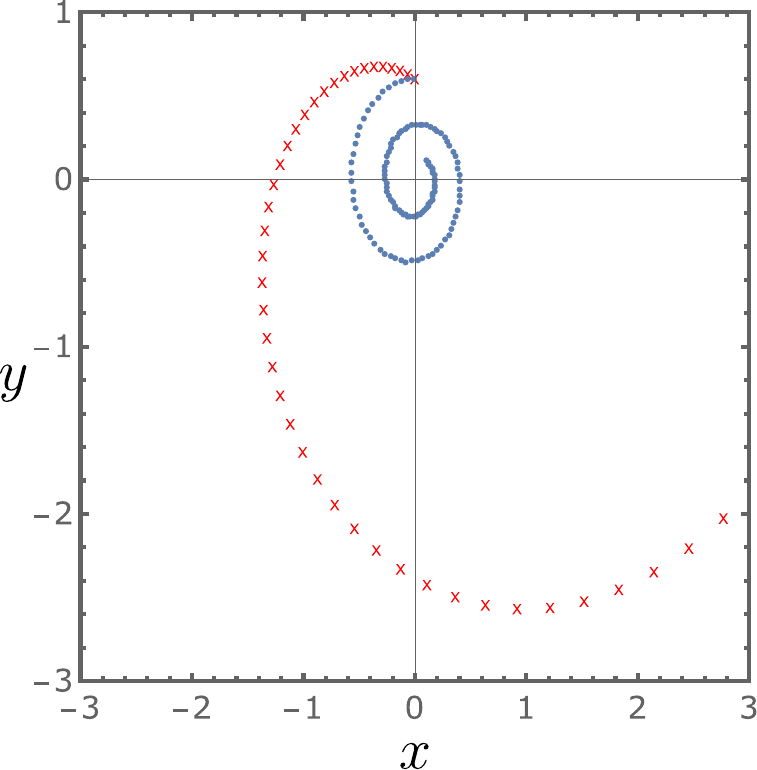}
	\caption{Evolution of the mean of the distribution function as derived through emulated integration by the Schr\"{o}dingerisation method with reduced resolution of the Fourier register ($\eta\in[-10,10]$ and $\Delta \eta = 0.1$). The quantum output (red crosses) forms an outward spiral and diverges from the classical output (blue dots).}\label{fig:graphicfokkerpl2demulschroedingerisationanddirectmatrexpo2}
\end{figure*}

{In Fig.~\ref{fig:graphicfokkerpl2demulschroedingerisationanddirectmatrexpo2}, we display the quantum output for a Fourier register discretised by $\Delta\eta = 0.1$. The quantum output (red crosses) diverges from the actual solution and forms an outward spiral. This demonstrates that by dreasing the resultion of the Fourier register, the quantum output can change qualitatively.}


\begin{thebibliography}{24}%
\makeatletter
\providecommand \@ifxundefined [1]{%
 \@ifx{#1\undefined}
}%
\providecommand \@ifnum [1]{%
 \ifnum #1\expandafter \@firstoftwo
 \else \expandafter \@secondoftwo
 \fi
}%
\providecommand \@ifx [1]{%
 \ifx #1\expandafter \@firstoftwo
 \else \expandafter \@secondoftwo
 \fi
}%
\providecommand \natexlab [1]{#1}%
\providecommand \enquote  [1]{``#1''}%
\providecommand \bibnamefont  [1]{#1}%
\providecommand \bibfnamefont [1]{#1}%
\providecommand \citenamefont [1]{#1}%
\providecommand \href@noop [0]{\@secondoftwo}%
\providecommand \href [0]{\begingroup \@sanitize@url \@href}%
\providecommand \@href[1]{\@@startlink{#1}\@@href}%
\providecommand \@@href[1]{\endgroup#1\@@endlink}%
\providecommand \@sanitize@url [0]{\catcode `\\12\catcode `\$12\catcode
  `\&12\catcode `\#12\catcode `\^12\catcode `\_12\catcode `\%12\relax}%
\providecommand \@@startlink[1]{}%
\providecommand \@@endlink[0]{}%
\providecommand \url  [0]{\begingroup\@sanitize@url \@url }%
\providecommand \@url [1]{\endgroup\@href {#1}{\urlprefix }}%
\providecommand \urlprefix  [0]{URL }%
\providecommand \Eprint [0]{\href }%
\providecommand \doibase [0]{http://dx.doi.org/}%
\providecommand \selectlanguage [0]{\@gobble}%
\providecommand \bibinfo  [0]{\@secondoftwo}%
\providecommand \bibfield  [0]{\@secondoftwo}%
\providecommand \translation [1]{[#1]}%
\providecommand \BibitemOpen [0]{}%
\providecommand \bibitemStop [0]{}%
\providecommand \bibitemNoStop [0]{.\EOS\space}%
\providecommand \EOS [0]{\spacefactor3000\relax}%
\providecommand \BibitemShut  [1]{\csname bibitem#1\endcsname}%
\let\auto@bib@innerbib\@empty
\bibitem [{\citenamefont {Jordan}\ and\ \citenamefont
  {Smith}(2007)}]{Jordan2007_nonlinear}%
  \BibitemOpen
  \bibfield  {author} {\bibinfo {author} {\bibfnamefont {D.}~\bibnamefont
  {Jordan}}\ and\ \bibinfo {author} {\bibfnamefont {P.}~\bibnamefont {Smith}},\
  }\href {https://books.google.co.uk/books?id=FmIS_0UEAUAC} {\emph {\bibinfo
  {title} {Nonlinear Ordinary Differential Equations: An Introduction for
  Scientists and Engineers}}},\ Oxford Texts in Applied and Engineering
  Mathematics\ (\bibinfo  {publisher} {OUP Oxford},\ \bibinfo {year}
  {2007})\BibitemShut {NoStop}%
\bibitem [{\citenamefont {Atkinson}\ \emph {et~al.}(2009)\citenamefont
  {Atkinson}, \citenamefont {Han},\ and\ \citenamefont
  {Stewart}}]{Atkinson2009_NumericalSolutionDiffEqs}%
  \BibitemOpen
  \bibfield  {author} {\bibinfo {author} {\bibfnamefont {K.~E.}\ \bibnamefont
  {Atkinson}}, \bibinfo {author} {\bibfnamefont {W.}~\bibnamefont {Han}}, \
  and\ \bibinfo {author} {\bibfnamefont {D.}~\bibnamefont {Stewart}},\ }\href
  {https://onlinelibrary.wiley.com/doi/abs/10.1002/9781118164495.ch7} {\emph
  {\bibinfo {title} {Numerical Solution of Ordinary Differential Equations}}}\
  (\bibinfo  {publisher} {John Wiley \& Sons, Ltd},\ \bibinfo {year}
  {2009})\BibitemShut {NoStop}%
\bibitem [{\citenamefont {Lubasch}\ \emph {et~al.}(2020)\citenamefont
  {Lubasch}, \citenamefont {Joo}, \citenamefont {Moinier}, \citenamefont
  {Kiffner},\ and\ \citenamefont {Jaksch}}]{Lubasch2020}%
  \BibitemOpen
  \bibfield  {author} {\bibinfo {author} {\bibfnamefont {M.}~\bibnamefont
  {Lubasch}}, \bibinfo {author} {\bibfnamefont {J.}~\bibnamefont {Joo}},
  \bibinfo {author} {\bibfnamefont {P.}~\bibnamefont {Moinier}}, \bibinfo
  {author} {\bibfnamefont {M.}~\bibnamefont {Kiffner}}, \ and\ \bibinfo
  {author} {\bibfnamefont {D.}~\bibnamefont {Jaksch}},\ }\href@noop {}
  {\bibfield  {journal} {\bibinfo  {journal} {Phys. Rev. A}\ }\textbf {\bibinfo
  {volume} {101}},\ \bibinfo {pages} {010301} (\bibinfo {year}
  {2020})}\BibitemShut {NoStop}%
\bibitem [{\citenamefont {Kyriienko}\ \emph {et~al.}(2021)\citenamefont
  {Kyriienko}, \citenamefont {Paine},\ and\ \citenamefont
  {Elfving}}]{Kyriienko2021}%
  \BibitemOpen
  \bibfield  {author} {\bibinfo {author} {\bibfnamefont {O.}~\bibnamefont
  {Kyriienko}}, \bibinfo {author} {\bibfnamefont {A.~E.}\ \bibnamefont
  {Paine}}, \ and\ \bibinfo {author} {\bibfnamefont {V.~E.}\ \bibnamefont
  {Elfving}},\ }\href@noop {} {\bibfield  {journal} {\bibinfo  {journal} {Phys.
  Rev. A}\ }\textbf {\bibinfo {volume} {103}},\ \bibinfo {pages} {052416}
  (\bibinfo {year} {2021})}\BibitemShut {NoStop}%
\bibitem [{\citenamefont {Shukla}\ and\ \citenamefont
  {Vedula}(2023)}]{Shukla2023}%
  \BibitemOpen
  \bibfield  {author} {\bibinfo {author} {\bibfnamefont {A.}~\bibnamefont
  {Shukla}}\ and\ \bibinfo {author} {\bibfnamefont {P.}~\bibnamefont
  {Vedula}},\ }\href@noop {} {\bibfield  {journal} {\bibinfo  {journal} {Appl.
  Math. and Comput.}\ }\textbf {\bibinfo {volume} {442}},\ \bibinfo {pages}
  {127708} (\bibinfo {year} {2023})}\BibitemShut {NoStop}%
\bibitem [{\citenamefont {Lloyd}\ \emph {et~al.}(2020)\citenamefont {Lloyd},
  \citenamefont {de~Palma}, \citenamefont {Gokler}, \citenamefont {Kiani},
  \citenamefont {Liu}, \citenamefont {Marvian}, \citenamefont {Tennie},\ and\
  \citenamefont {Palmer}}]{Lloyd2020}%
  \BibitemOpen
  \bibfield  {author} {\bibinfo {author} {\bibfnamefont {S.}~\bibnamefont
  {Lloyd}}, \bibinfo {author} {\bibfnamefont {G.}~\bibnamefont {de~Palma}},
  \bibinfo {author} {\bibfnamefont {C.}~\bibnamefont {Gokler}}, \bibinfo
  {author} {\bibfnamefont {B.}~\bibnamefont {Kiani}}, \bibinfo {author}
  {\bibfnamefont {Z.-W.}\ \bibnamefont {Liu}}, \bibinfo {author} {\bibfnamefont
  {M.}~\bibnamefont {Marvian}}, \bibinfo {author} {\bibfnamefont
  {F.}~\bibnamefont {Tennie}}, \ and\ \bibinfo {author} {\bibfnamefont
  {T.}~\bibnamefont {Palmer}},\ }\href@noop {} {\bibfield  {journal} {\bibinfo
  {journal} {arXiv}\ }\textbf {\bibinfo {volume} {2011.06571}} (\bibinfo {year}
  {2020})}\BibitemShut {NoStop}%
\bibitem [{\citenamefont {Liu}\ \emph {et~al.}(2021)\citenamefont {Liu},
  \citenamefont {Kolden}, \citenamefont {Krovi}, \citenamefont {Loureiro},
  \citenamefont {Trivisa},\ and\ \citenamefont {Childs}}]{Liu2021}%
  \BibitemOpen
  \bibfield  {author} {\bibinfo {author} {\bibfnamefont {J.-P.}\ \bibnamefont
  {Liu}}, \bibinfo {author} {\bibfnamefont {H.~O.}\ \bibnamefont {Kolden}},
  \bibinfo {author} {\bibfnamefont {H.~K.}\ \bibnamefont {Krovi}}, \bibinfo
  {author} {\bibfnamefont {N.}~\bibnamefont {Loureiro}}, \bibinfo {author}
  {\bibfnamefont {K.}~\bibnamefont {Trivisa}}, \ and\ \bibinfo {author}
  {\bibfnamefont {A.}~\bibnamefont {Childs}},\ }\href@noop {} {\bibfield
  {journal} {\bibinfo  {journal} {PNAS}\ }\textbf {\bibinfo {volume} {118}},\
  \bibinfo {pages} {2026805118} (\bibinfo {year} {2021})}\BibitemShut {NoStop}%
\bibitem [{\citenamefont {Krovi}(2023)}]{Krovi2023}%
  \BibitemOpen
  \bibfield  {author} {\bibinfo {author} {\bibfnamefont {H.}~\bibnamefont
  {Krovi}},\ }\href@noop {} {\bibfield  {journal} {\bibinfo  {journal}
  {{Quantum}}\ }\textbf {\bibinfo {volume} {7}},\ \bibinfo {pages} {913}
  (\bibinfo {year} {2023})}\BibitemShut {NoStop}%
\bibitem [{\citenamefont {Risken}(2012)}]{risken2012fokker}%
  \BibitemOpen
  \bibfield  {author} {\bibinfo {author} {\bibfnamefont {H.}~\bibnamefont
  {Risken}},\ }\href {https://books.google.co.uk/books?id=dXvpCAAAQBAJ} {\emph
  {\bibinfo {title} {The Fokker-Planck Equation: Methods of Solution and
  Applications}}},\ Springer Series in Synergetics\ (\bibinfo  {publisher}
  {Springer Berlin Heidelberg},\ \bibinfo {year} {2012})\BibitemShut {NoStop}%
\bibitem [{\citenamefont {Pawula}(1967)}]{Pawula1967}%
  \BibitemOpen
  \bibfield  {author} {\bibinfo {author} {\bibfnamefont {R.~F.}\ \bibnamefont
  {Pawula}},\ }\href {\doibase 10.1103/PhysRev.162.186} {\bibfield  {journal}
  {\bibinfo  {journal} {Phys. Rev.}\ }\textbf {\bibinfo {volume} {162}},\
  \bibinfo {pages} {186} (\bibinfo {year} {1967})}\BibitemShut {NoStop}%
\bibitem [{\citenamefont {Childs}\ \emph {et~al.}(2021)\citenamefont {Childs},
  \citenamefont {Liu},\ and\ \citenamefont {Ostrander}}]{Childs2021}%
  \BibitemOpen
  \bibfield  {author} {\bibinfo {author} {\bibfnamefont {A.~M.}\ \bibnamefont
  {Childs}}, \bibinfo {author} {\bibfnamefont {J.-P.}\ \bibnamefont {Liu}}, \
  and\ \bibinfo {author} {\bibfnamefont {A.}~\bibnamefont {Ostrander}},\ }\href
  {\doibase 10.22331/q-2021-11-10-574} {\bibfield  {journal} {\bibinfo
  {journal} {{Quantum}}\ }\textbf {\bibinfo {volume} {5}},\ \bibinfo {pages}
  {574} (\bibinfo {year} {2021})}\BibitemShut {NoStop}%
\bibitem [{\citenamefont {Harrow}\ \emph {et~al.}(2009)\citenamefont {Harrow},
  \citenamefont {Hassidim},\ and\ \citenamefont {Lloyd}}]{Harrow2009}%
  \BibitemOpen
  \bibfield  {author} {\bibinfo {author} {\bibfnamefont {A.~W.}\ \bibnamefont
  {Harrow}}, \bibinfo {author} {\bibfnamefont {A.}~\bibnamefont {Hassidim}}, \
  and\ \bibinfo {author} {\bibfnamefont {S.}~\bibnamefont {Lloyd}},\ }\href
  {\doibase 10.1103/PhysRevLett.103.150502} {\bibfield  {journal} {\bibinfo
  {journal} {Phys. Rev. Lett.}\ }\textbf {\bibinfo {volume} {103}},\ \bibinfo
  {pages} {150502} (\bibinfo {year} {2009})}\BibitemShut {NoStop}%
\bibitem [{\citenamefont {Childs}\ \emph {et~al.}(2017)\citenamefont {Childs},
  \citenamefont {Kothari},\ and\ \citenamefont {Somma}}]{Childs2017}%
  \BibitemOpen
  \bibfield  {author} {\bibinfo {author} {\bibfnamefont {A.~M.}\ \bibnamefont
  {Childs}}, \bibinfo {author} {\bibfnamefont {R.}~\bibnamefont {Kothari}}, \
  and\ \bibinfo {author} {\bibfnamefont {R.~D.}\ \bibnamefont {Somma}},\ }\href
  {\doibase 10.1137/16M1087072} {\bibfield  {journal} {\bibinfo  {journal}
  {SIAM J. Comput.}\ }\textbf {\bibinfo {volume} {46}},\ \bibinfo {pages}
  {1920} (\bibinfo {year} {2017})},\ \Eprint
  {http://arxiv.org/abs/https://doi.org/10.1137/16M1087072}
  {https://doi.org/10.1137/16M1087072} \BibitemShut {NoStop}%
\bibitem [{\citenamefont {Holubec}\ \emph {et~al.}(2019)\citenamefont
  {Holubec}, \citenamefont {Kroy},\ and\ \citenamefont
  {Steffenoni}}]{Holubec2019}%
  \BibitemOpen
  \bibfield  {author} {\bibinfo {author} {\bibfnamefont {V.}~\bibnamefont
  {Holubec}}, \bibinfo {author} {\bibfnamefont {K.}~\bibnamefont {Kroy}}, \
  and\ \bibinfo {author} {\bibfnamefont {S.}~\bibnamefont {Steffenoni}},\
  }\href {\doibase 10.1103/PhysRevE.99.032117} {\bibfield  {journal} {\bibinfo
  {journal} {Phys. Rev. E}\ }\textbf {\bibinfo {volume} {99}},\ \bibinfo
  {pages} {032117} (\bibinfo {year} {2019})}\BibitemShut {NoStop}%
\bibitem [{\citenamefont {Berry}(2014)}]{Berry2014}%
  \BibitemOpen
  \bibfield  {author} {\bibinfo {author} {\bibfnamefont {D.~W.}\ \bibnamefont
  {Berry}},\ }\href {\doibase 10.1088/1751-8113/47/10/105301} {\bibfield
  {journal} {\bibinfo  {journal} {J. Phys. A}\ }\textbf {\bibinfo {volume}
  {47}},\ \bibinfo {pages} {105301} (\bibinfo {year} {2014})}\BibitemShut
  {NoStop}%
\bibitem [{\citenamefont {Berry}\ \emph {et~al.}(2017)\citenamefont {Berry},
  \citenamefont {Childs}, \citenamefont {Ostrander},\ and\ \citenamefont
  {Wang}}]{Berry2017}%
  \BibitemOpen
  \bibfield  {author} {\bibinfo {author} {\bibfnamefont {D.~W.}\ \bibnamefont
  {Berry}}, \bibinfo {author} {\bibfnamefont {A.~M.}\ \bibnamefont {Childs}},
  \bibinfo {author} {\bibfnamefont {A.}~\bibnamefont {Ostrander}}, \ and\
  \bibinfo {author} {\bibfnamefont {G.}~\bibnamefont {Wang}},\ }\href@noop {}
  {\bibfield  {journal} {\bibinfo  {journal} {Commun. Math. Phys.}\ }\textbf
  {\bibinfo {volume} {356}},\ \bibinfo {pages} {1057} (\bibinfo {year}
  {2017})}\BibitemShut {NoStop}%
\bibitem [{\citenamefont {Childs}\ and\ \citenamefont
  {Liu}(2020)}]{Childs2020}%
  \BibitemOpen
  \bibfield  {author} {\bibinfo {author} {\bibfnamefont {A.~M.}\ \bibnamefont
  {Childs}}\ and\ \bibinfo {author} {\bibfnamefont {J.~P.}\ \bibnamefont
  {Liu}},\ }\href@noop {} {\bibfield  {journal} {\bibinfo  {journal} {Commun.
  Math. Phys.}\ }\textbf {\bibinfo {volume} {375}},\ \bibinfo {pages}
  {1427–1457} (\bibinfo {year} {2020})}\BibitemShut {NoStop}%
\bibitem [{\citenamefont {Vedral}\ \emph {et~al.}(1996)\citenamefont {Vedral},
  \citenamefont {Barenco},\ and\ \citenamefont {Ekert}}]{Vedral1996}%
  \BibitemOpen
  \bibfield  {author} {\bibinfo {author} {\bibfnamefont {V.}~\bibnamefont
  {Vedral}}, \bibinfo {author} {\bibfnamefont {A.}~\bibnamefont {Barenco}}, \
  and\ \bibinfo {author} {\bibfnamefont {A.}~\bibnamefont {Ekert}},\ }\href
  {\doibase 10.1103/PhysRevA.54.147} {\bibfield  {journal} {\bibinfo  {journal}
  {Phys. Rev. A}\ }\textbf {\bibinfo {volume} {54}},\ \bibinfo {pages} {147}
  (\bibinfo {year} {1996})}\BibitemShut {NoStop}%
\bibitem [{\citenamefont {An}\ \emph {et~al.}(2022)\citenamefont {An},
  \citenamefont {Liu}, \citenamefont {Wang},\ and\ \citenamefont
  {Zhao}}]{An2022}%
  \BibitemOpen
  \bibfield  {author} {\bibinfo {author} {\bibfnamefont {D.}~\bibnamefont
  {An}}, \bibinfo {author} {\bibfnamefont {J.~P.}\ \bibnamefont {Liu}},
  \bibinfo {author} {\bibfnamefont {D.}~\bibnamefont {Wang}}, \ and\ \bibinfo
  {author} {\bibfnamefont {Q.}~\bibnamefont {Zhao}},\ }\href@noop {} {\bibfield
   {journal} {\bibinfo  {journal} {arXiv}\ }\textbf {\bibinfo {volume}
  {2211.05246}} (\bibinfo {year} {2022})}\BibitemShut {NoStop}%
\bibitem [{\citenamefont {Lloyd}\ \emph {et~al.}(2021)\citenamefont {Lloyd},
  \citenamefont {Kiani}, \citenamefont {Arvidsson-Shukur}, \citenamefont
  {Bosch}, \citenamefont {De~Palma}, \citenamefont {Kaminsky}, \citenamefont
  {Liu},\ and\ \citenamefont {Marvian}}]{Lloyd2021}%
  \BibitemOpen
  \bibfield  {author} {\bibinfo {author} {\bibfnamefont {S.}~\bibnamefont
  {Lloyd}}, \bibinfo {author} {\bibfnamefont {B.~T.}\ \bibnamefont {Kiani}},
  \bibinfo {author} {\bibfnamefont {D.~R.~M.}\ \bibnamefont
  {Arvidsson-Shukur}}, \bibinfo {author} {\bibfnamefont {S.}~\bibnamefont
  {Bosch}}, \bibinfo {author} {\bibfnamefont {G.}~\bibnamefont {De~Palma}},
  \bibinfo {author} {\bibfnamefont {W.~M.}\ \bibnamefont {Kaminsky}}, \bibinfo
  {author} {\bibfnamefont {Z.-W.}\ \bibnamefont {Liu}}, \ and\ \bibinfo
  {author} {\bibfnamefont {M.}~\bibnamefont {Marvian}},\ }\href@noop {}
  {\bibfield  {journal} {\bibinfo  {journal} {arXiv}\ }\textbf {\bibinfo
  {volume} {2104.01410}} (\bibinfo {year} {2021})}\BibitemShut {NoStop}%
\bibitem [{\citenamefont {Jin}\ \emph {et~al.}(2023)\citenamefont {Jin},
  \citenamefont {Liu},\ and\ \citenamefont {Yu}}]{Jin2023}%
  \BibitemOpen
  \bibfield  {author} {\bibinfo {author} {\bibfnamefont {S.}~\bibnamefont
  {Jin}}, \bibinfo {author} {\bibfnamefont {N.}~\bibnamefont {Liu}}, \ and\
  \bibinfo {author} {\bibfnamefont {Y.}~\bibnamefont {Yu}},\ }\href {\doibase
  10.1103/PhysRevA.108.032603} {\bibfield  {journal} {\bibinfo  {journal}
  {Phys. Rev. A}\ }\textbf {\bibinfo {volume} {108}},\ \bibinfo {pages}
  {032603} (\bibinfo {year} {2023})}\BibitemShut {NoStop}%
\bibitem [{\citenamefont {Nielsen}\ and\ \citenamefont
  {Chuang}(2010)}]{Nielson2010}%
  \BibitemOpen
  \bibfield  {author} {\bibinfo {author} {\bibfnamefont {M.~A.}\ \bibnamefont
  {Nielsen}}\ and\ \bibinfo {author} {\bibfnamefont {I.~L.}\ \bibnamefont
  {Chuang}},\ }\href@noop {} {\emph {\bibinfo {title} {Quantum Computation and
  Quantum Information: 10th Anniversary Edition.}}}\ (\bibinfo  {publisher}
  {Cambridge University Press},\ \bibinfo {year} {2010})\BibitemShut {NoStop}%
\bibitem [{\citenamefont {Keizer}(1972)}]{Keitzer1972}%
  \BibitemOpen
  \bibfield  {author} {\bibinfo {author} {\bibfnamefont {J.}~\bibnamefont
  {Keizer}},\ }\href@noop {} {\bibfield  {journal} {\bibinfo  {journal} {J.
  Stat. Phys.}\ }\textbf {\bibinfo {volume} {6}},\ \bibinfo {pages} {67072}
  (\bibinfo {year} {1972})}\BibitemShut {NoStop}%
\bibitem [{\citenamefont {Camps}\ \emph {et~al.}(2022)\citenamefont {Camps},
  \citenamefont {Lin}, \citenamefont {Van~Beeumen},\ and\ \citenamefont
  {Yang}}]{Camps2022}%
  \BibitemOpen
  \bibfield  {author} {\bibinfo {author} {\bibfnamefont {D.}~\bibnamefont
  {Camps}}, \bibinfo {author} {\bibfnamefont {L.}~\bibnamefont {Lin}}, \bibinfo
  {author} {\bibfnamefont {R.}~\bibnamefont {Van~Beeumen}}, \ and\ \bibinfo
  {author} {\bibfnamefont {C.}~\bibnamefont {Yang}},\ }\href@noop {} {\bibfield
   {journal} {\bibinfo  {journal} {arXiv}\ }\textbf {\bibinfo {volume}
  {2203.10236}} (\bibinfo {year} {2022})}\BibitemShut {NoStop}%
\end{thebibliography}
\end{document}